\documentclass[a4paper,11pt]{article}
\pdfoutput=1 
\usepackage{jheppub} 
\usepackage{bbm,bm,graphicx,mathtools,color,hyperref,slashed}
\usepackage{amssymb,amsmath}
\usepackage[T1]{fontenc}
\usepackage[caption=false]{subfig}

\graphicspath{{./Figures/}}

\newcommand{\diff}{\mathrm{d}}
\newcommand{\p}{\partial}

\newcommand{\Diff}{{\mathcal{D}}}

\newcommand{\be}{\begin{equation}}      
\newcommand{\ee}{\end{equation}}      
\newcommand{\bea}{\begin{eqnarray}}      
\newcommand{\eea}{\end{eqnarray}}

\newcommand{\im}{\mathrm{i}}

\newcommand{\calA}{\mathcal{A}}
\newcommand{\calG}{\mathcal{G}}
\newcommand{\calP}{\mathcal{P}}

\title{Vacuum structure of bifundamental gauge theories at finite topological angles}

\author[a]{Yuya Tanizaki,}
\author[b,c]{Yuta Kikuchi}

\affiliation[a]{RIKEN BNL Research Center, Brookhaven National Laboratory, Upton, NY 11973 USA}
\affiliation[b]{Department of Physics and Astronomy, Stony Brook University, Stony Brook, New York 11794-3800, USA}
\affiliation[c]{Department of Physics, Kyoto University, Kyoto 606-8502, Japan}

\emailAdd{yuya.tanizaki@riken.jp}
\emailAdd{yuta.kikuchi@stonybrook.edu}

\abstract{
We discuss possible vacuum structures of $SU(n)\times SU(n)$ gauge theories with bifundamental matters at finite $\theta$ angles. 
In order to give a precise constraint, a mixed 't~Hooft anomaly is studied in detail by gauging the center $\mathbb{Z}_n$ one-form symmetry of the bifundamental gauge theory. 
We propose phase diagrams that are consistent with the constraints, and also give a heuristic explanation of the result based on the dual superconductor scenario of confinement. 
}

\begin{document}
\maketitle
\section{Introduction}\label{sec:introduction}

Dynamics of non-Abelian gauge theories depends not only on the gauge coupling constant but also on the topological $\theta$ angle. Since its discovery, the dependence of vacua and excitations on the parameter $\theta$ has been a key issue to understand the topological nature of gauge theories~\cite{Polyakov:1975rs, Belavin:1975fg, tHooft:1976rip, Callan:1976je, Jackiw:1976pf, Coleman:1976uz}. 
The strongly interacting sector in the Standard Model of particle physics is the $SU(3)$ vector-like gauge theory, and thus all the interactions preserve the $CP$ invariance except for this topological term.  
It is widely believed that the $\theta$ angle of quantum chromodynamics (QCD) is quite small because $CP$ in the strong sector is well maintained in our universe according to the experiment on neutron's electric dipole moment~\cite{Baker:2006ts}. 

For four-dimensional $SU(n)$ Yang-Mills theory, the angle $\theta$ is periodic in $2\pi$, and thus the requirement of $CP$ invariance of theory raises two candidates: $\theta=0$ and $\theta=\pi$. 
Understandings on the vacuum structure at $\theta=\pi$ are of particular importance, and many studies have been devoted to it using various techniques, including large-$n$ limit, effective models, and chiral perturbation~\cite{Witten:1980sp, tHooft:1981bkw, Ohta:1981ai, Cardy:1981qy, Cardy:1981fd, Wiese:1988qz, Affleck:1991tj, Creutz:1995wf, Creutz:2009kx, Smilga:1998dh, Witten:1998uka, Halperin:1998rc, Boer:2008ct, Aoki:2014moa, Mameda:2014cxa, Verbaarschot:2014upa}. 
In certain limits (for example large $n$), one can show that $SU(n)$ Yang-Mills theory possesses the first-order phase transition at $\theta=\pi$ and breaks $CP$ spontaneously. 
This tells us that physics at $\theta=\pi$ is dramatically different from that at $\theta=0$, and it is not known what would happen in generic cases. 
Recently, in Ref.~\cite{Gaiotto:2017yup}, a new technique has been developed for $SU(n)$ pure Yang-Mills theory and also for $SU(n)$ Yang-Mills theory with adjoint matter fields, which gives a rigorous constraint on the vacuum structure at $\theta=\pi$ by discussing an 't Hooft anomaly matching. 
More interestingly, it reveals under reasonable assumptions that the first-order phase transition at $\theta=\pi$ survives at finite temperatures at least until the deconfinement transition happens. 
The purpose of this paper is to extend and apply their technique to study the $\theta=\pi$ dynamics of other gauge theories, especially $SU(n)\times SU(n)$ Yang-Mills theory with bifundamental matter fields. 

Bifundamental gauge theories have acquired an interesting position among a lot of gauge theories. 
The number of color $n$ provides a hidden expansion parameter of strongly coupled gauge theories~\cite{tHooft:1973alw}, and the limit $n\to\infty$ is governed by the planar diagrams. 
The same limit has also an interesting possibility to relate non-supersymmetric gauge theories and supersymmetric Yang-Mills theory~\cite{Strassler:2001fs, Kachru:1998ys, Lawrence:1998ja, Bershadsky:1998mb, Bershadsky:1998cb, Kakushadze:1998tr, Kakushadze:1998tz, Schmaltz:1998bg, Armoni:1999gc}. 
Let us pick up an $SU(n)\times SU(n)$ gauge theory with one bifundamental Dirac fermion for example. 
Although it is not supersymmetric, it is a daughter theory of the orbifold equivalence to $\mathcal{N}=1$ supersymmetric $SU(2n)$ Yang-Mills theory at least diagrammatically. The condition of the nonperturbative orbifold equivalence has also been discussed extensively, and one must know the vacuum structure to judge the equivalence~\cite{Armoni:2003gp, Armoni:2003fb, Armoni:2003yv, Armoni:2004uu, Armoni:2004ub, Kovtun:2003hr, Kovtun:2004bz, Kovtun:2005kh, Armoni:2005wta}. 
Although it is not yet known whether the nonperturbative equivalence holds for bifundamental gauge theories, they exhibit rich dynamics~\cite{Kovtun:2004bz, Kovtun:2005kh, Armoni:2005wta} and it is an important and interesting topic to study them in order to deepen our understandings on nonperturbative gluon-dynamics. 
The supersymmetric bifundamental gauge theory is also an interesting topic for the same purpose, and the string theory picture is very useful there~\cite{Amariti:2016hlj}. 

In this paper, we give a rigorous constraint on the vacuum structure of $SU(n)\times SU(n)$ gauge theories with bifundamental matter fields at finite topological angles. Since the theory has two $SU(n)$ gauge groups, it has two topological angles $\theta_1$ and $\theta_2$. 
The theory is $CP$ invariant at $(\theta_1,\theta_2)=(0,0)$, $(\pi,0)$, $(0,\pi)$ and $(\pi,\pi)$, and we discuss the global consistency of 't Hooft anomalies to see whether the vacuum is continuously connected without breaking $CP$ at those points. 
We propose phase diagrams in the $\theta_1$-$\theta_2$ plane that are consistent with the constraints, and give its heuristic interpretation based on the dual superconductor model of confinement. 

This paper is organized as follows: In Section~\ref{sec:review}, we review the basics for the $SU(n)$ Yang-Mills theory to make the paper self-contained. We also give a review on how the rigorous constraint on the $CP$ symmetry at $\theta=\pi$ can be derived for it. 
In Section~\ref{sec:SU(n)2_bifundamental}, we discuss the bifundamental $SU(n)\times SU(n)$ gauge theory at finite topological angles, and interpret our result based on the dual superconductor model of confinement. 
We give conclusions in Section~\ref{sec:conclusion}. We give a review on necessary computations of topological field theories in Appendix~\ref{sec:tft}. 

\section{Review on $\mathfrak{su}(n)$ Yang-Mills theory}\label{sec:review}

In this section, we give a brief review on four-dimensional Yang-Mills theory with the gauge Lie algebra $\mathfrak{su}(n)$ in order to make the paper self-contained. Especially, we consider the case  where the gauge group is $SU(n)$ or $PSU(n)=SU(n)/\mathbb{Z}_n$. For more details of this subject, see e.g. Refs.~\cite{Aharony:2013hda, Kapustin:2014gua, Tachikawa:2014mna, Gaiotto:2014kfa}. 

\subsection{Electric and magnetic charges}

We first discuss all the possible electric and magnetic charges especially for the gauge group $G=SU(n)$ and $G=SU(n)/\mathbb{Z}_n$. For slightly more general settings, let $\widehat{G}=SU(n)$ be the universal cover of the gauge group, and the gauge group is given by $G=\widehat{G}/H$ with a center subgroup $H<\mathbb{Z}_n$. 

Let us consider the electric charge first. Classically, all the electric charges belong to representations of the Lie algebra $\mathfrak{su}(n)$, i.e., elements of the weight lattice. After quantization, each electric charge can emit and absorb gluons that belong to the adjoint representations, and thus only the number of boxes in Young tableau (mod $n$) are relevant to characterize the electric charge for low-energy dynamics. 
We can then label the electric charge of (test) particles by $z_e\in\mathbb{Z}_n$ for any $\mathfrak{su}(n)$ gauge theories to discuss infrared properties. 
When the gauge group is $G=SU(n)/H$, the particle must be invariant under $H<\mathbb{Z}_n$ and only such $z_e\in \mathbb{Z}_n$ are allowed. When $G=SU(n)$, the allowed electric charges are $z_e=0,1,\ldots,n-1$. For $G=SU(n)/\mathbb{Z}_n$, the only allowed electric charge is $z_e=0$. 
The electric charge of dynamical particles must be some of these charges, too, but not all of them need to be dynamical. 

Magnetic charges are in the representation of the GNO dual gauge group $G^{\vee}$~\cite{Goddard:1976qe}. Universal covers of the original and dual gauge groups $\widehat{G}$, $\widehat{G^{\vee}}$ have the same center group, and thus electric and magnetic charges for candidates of test particles are labeled by 
\be
(z_e,z_m)\in\mathbb{Z}_n\times \mathbb{Z}_n. 
\ee
For test particles being genuine point-like objects, the set of allowed charges must satisfy Dirac quantization condition: For both $(z_e,z_m)$ and $(z'_e,z'_m)$ being test particles, they must satisfy  
\be
{1\over n}(z_e z'_m-z'_e z_m)=0 \quad \mathrm{mod}\; 1. 
\ee
This is also called the mutual locality condition. 

\subsection{$SU(n)$ Yang-Mills theory and its genuine line operators}
The four-dimensional $SU(n)$ pure Yang-Mills theory is described by, 
\be
S=-{1\over 2g^2}\int\mathrm{Tr} (G\wedge *G)+{\im\theta\over 8\pi^2}\int \mathrm{Tr}(G\wedge G),  
\label{eq:SU(n)YM}
\ee
where $G$ is the field strength of the $SU(n)$ gauge field $a$:
\be
G=\diff a+a\wedge a. 
\ee
In our convention, $a=\im a_{i\mu}T^i\diff x^\mu$ is locally an $n\times n$ anti-Hermitian matrix-valued one-form, and $\mathrm{Tr}(T^i T^j)={1\over 2}\delta^{ij}$. 
The theory is invariant under the $SU(n)$ gauge transformation $a\mapsto g^{-1}a g+g^{-1}\diff g$, and the physical observables must respect the gauge invariance. 
The Wilson line in the fundamental representation along a closed line $C$ is a gauge invariant object, 
\be
W(C)=\mathrm{Tr}\left[\mathcal{P}\exp\oint_C a\right]. 
\label{eq:SU(n)Wilson}
\ee
One can measure the electric charge $z_e=1$ of the Wilson line by introducing a topological surface operator~\cite{Gaiotto:2014kfa, Gukov:2008sn}. In this sense, $SU(n)$ Yang-Mills theory has a global center symmetry that is called electric $\mathbb{Z}_n$ one-form symmetry. 

Since theory has a fundamental Wilson line in the spectrum of genuine line operators as mentioned above, there is no magnetic line operators as a genuine line object. Indeed, let $(z_e,z_m)$ be a charge of the line operator, then the Dirac quantization condition with the fundamental Wilson line with charge $(1,0)$ claims 
\be
z_m=0\quad \mathrm{mod}\; n.
\ee
This means that there is no magnetic or dyonic genuine line. The genuine line operators with different electric charges are given by $W(C)^k$ with $z_e=k=0,1,\ldots, n-1$. 

\subsection{$SU(n)/\mathbb{Z}_n$ Yang-Mills theory}\label{sec:PSU(n)YM}

Let us next consider the $SU(n)/\mathbb{Z}_n$ gauge theory, and the general argument on the electric charge shows that the purely electric line operators must be invariant under $\mathbb{Z}_n$, such as $W(C)^{n}$. 
Since the Dirac quantization condition with allowed electric particles does not give any constraints on $z_m$, the genuine line with $z_m=1$ is possible. 
Let us assume that we have a theory with a magnetic or dyonic line with charge $(z_e,z_m)=(-p,1)$ with some $p=0,1,\ldots, n-1$. The Dirac quantization says that the charge $(z'_e,z'_m)$ of other genuine line operators must satisfy 
\be
z'_e=-p z'_m\quad \mathrm{mod}\;n. 
\ee
Therefore, the electric charge of line operators with $z_m=1$ is fixed to $-p$ once the line with $(z_e,z_m)=(-p,1)$ exists.  $p$ is a new parameter of $SU(n)/\mathbb{Z}_n$ gauge theories, which is called the discrete theta angle~\cite{Aharony:2013hda, Kapustin:2014gua, Tachikawa:2014mna, Gaiotto:2014kfa}, and it specifies the spectrum of genuine line operators. 

We can construct $SU(n)/\mathbb{Z}_n$ Yang-Mills theory by coupling $SU(n)$ Yang-Mills theory (\ref{eq:SU(n)YM}) to the following $\mathbb{Z}_n$ topological field theory~\cite{Kapustin:2014gua}, 
\be
S_{\mathrm{TFT}}={\im\over 2\pi}\int F\wedge (\diff A+n B)+{\im np\over 4\pi }\int B\wedge B. 
\label{eq:ZnTFT}
\ee
This topological field theory is a low-energy effective description of the spontaneous (one-form) gauge symmetry breaking $U(1)\to \mathbb{Z}_n$ when the fields with charge $n$ are condensed~\cite{Banks:2010zn}. 
Here, $A$ and $B$ are one-form and two-form $U(1)$ gauge fields, respectively, and $F$ is a two-form auxiliary field (see Appendix~\ref{sec:tft} for this topological field theory). We require that the action is invariant under the one-form $U(1)$ gauge transformation, 
\be
A\mapsto A-n\lambda, \; B\mapsto B+\diff \lambda, \; F\mapsto F-p\diff \lambda, 
\label{eq:U1one_form_gauge}
\ee
and then $p$ must be an integer due to the gauge invariance\footnote{In this paper, we implicitly assume that we consider field theories only on spin manifolds}. 

In order to couple $SU(n)$ Yang-Mills theory to $\mathbb{Z}_n$ topological field theory (\ref{eq:ZnTFT}), we first extend the gauge group from $SU(n)$ to $U(n)=(SU(n)\times U(1))/\mathbb{Z}_n$, and identify this $U(1)$ factor with that of the $U(1)$ gauge field $A$ in (\ref{eq:ZnTFT}). 
Correspondingly, the $SU(n)$ gauge field $a$ is replaced by the $U(n)$ gauge field, 
\be
\calA=a+{1\over n}A\bm{1}_n, 
\ee
and the gauge field strength becomes 
\be
\calG=\diff \calA+\calA\wedge \calA. 
\ee
Under the $U(1)$ one-form gauge transformation, $\calG$ is transformed as 
\be
\calG\mapsto \calG-\diff \lambda\bm{1}_n. 
\ee
In order to obtain the $SU(n)/\mathbb{Z}_n$ gauge theory instead of $U(n)$ gauge  theory, we postulate the invariance under the $U(1)$ one-form gauge transformation, and then the gauge invariant combination is given by $\calG+ B\bm{1}_n$ (for notational simplicity, the identity matrix $\bm{1}_n$ will be omitted below). 
As a result, the classical action for the $SU(n)/\mathbb{Z}_n$ Yang-Mills theory is given by
\bea
S&=&-{1\over 2g^2}\int\mathrm{Tr} ((\calG+B)\wedge *(\calG+B))+{\im\theta\over 8\pi^2}\int \mathrm{Tr}((\calG+B)\wedge (\calG+B))\nonumber\\
&&+{\im\over 2\pi}\int F\wedge (\diff A+nB)+{\im np\over 4\pi }\int B\wedge B. 
\label{eq:PSU(n)YM}
\eea
Locally, we obtain $B=-{1\over n}\diff A$ by integrating out $F$, and its substitution recovers the original $SU(n)$ Yang-Mills action but this operation is ill-defined globally. Spectrum of local operators on topologically trivial manifolds is unchanged by this gauging procedure, but there is a crucial difference on nontrivial topologies or with non-local operators as we shall see below. 

If one tries to define the Wilson line by (\ref{eq:SU(n)Wilson}), it is not gauge invariant under the $U(n)$ gauge transformation. 
We can define two kinds of gauge-invariant line operators with $(z_e,z_m)=(1,0)$ and $(0,1)$ but they need a topological surface ($\p \Sigma=C$) in general in order to maintain the $U(n)$ $0$-form and $U(1)$ $1$-form gauge invariance:
\bea
W(C,\Sigma)&=&\mathrm{Tr}\left[\mathcal{P}\exp\oint_C a\right]\exp\left[{1\over n}\oint_C A+\int_\Sigma B\right],\\
H(C,\Sigma)&=&\exp\left[\int_\Sigma(F+pB)\right]. 
\eea
We now claim that the action (\ref{eq:PSU(n)YM}) indeed describes the Yang-Mills theory of gauge group $SU(n)/\mathbb{Z}_n$ with the discrete theta angle $p$. Indeed, let us consider $H(C,\Sigma)W(C,\Sigma)^{-p}$ that has charge $(z_e,z_m)=(-p,1)$:
\be
H(C,\Sigma)W(C,\Sigma)^{-p}=\exp\left(\int_\Sigma F\right)\left(\mathrm{Tr}\left[\mathcal{P}\exp\oint_C a\right]\exp\left[{1\over n}\oint_C A\right]\right)^{-p}. 
\ee
Superficially, it depends on the surface $\Sigma$, but the equation of motion of $A$ claims that ${1\over 2\pi\im}F\in H^2(X,\mathbb{Z})$, and thus $\exp(\int_\Sigma F)$ does not depend on the choice of surfaces $\Sigma$ satisfying $\p \Sigma=C$.  
Therefore, the theory (\ref{eq:PSU(n)YM}) have the dyonic genuine line operator with charge $(z_e,z_m)=(-p,1)$, which is concretely given by $H(C,\Sigma)W(C,\Sigma)^{-p}$. 

Let us also explain why $p$ is called the discrete theta angle. For this purpose, we consider the shift $\theta\mapsto \theta+2\pi$.  The change of the action (\ref{eq:PSU(n)YM}) under this shift is given by
\be
\Delta S={\im\over 4\pi}\int \mathrm{Tr}((\calG+B)\wedge (\calG+B))={\im\over 4\pi}\int\mathrm{Tr}(\calG\wedge \calG)-{\im n\over 4\pi}\int B\wedge B. 
\ee
The first term is in $2\pi\im\mathbb{Z}$ on spin four-manifolds due to the index theorem, and thus the $2\pi$ shift of $\theta$ changes $p$ to $p-1$ (mod $n$). This is a consequence of the fact that the electric charge of dyons is shifted by $\theta/2\pi$ because of the $\theta$ angle, and often called the Witten effect~\cite{Witten:1979ey}. As a result, the periodicity of $\theta$ is extended to $2\pi n$ from $2\pi$. 
Since $n$ different choices of $p$ for the $SU(n)/\mathbb{Z}_n$ gauge theory is related by $2\pi$ shifts of $\theta$, $p$ is called the discrete theta angle, although this is not always true for other gauge groups~\cite{Aharony:2013hda}.

\subsection{Spontaneous $CP$ breaking at $\theta=\pi$ of $SU(n)$ Yang-Mills theory}\label{sec:CP_SU(n)}

We also review how one can claim the spontaneous breaking of $CP$ at $\theta=\pi$ following the procedure with use of an 't Hooft anomaly, which was recently developed in Ref.~\cite{Gaiotto:2017yup}.  
We assume that $SU(n)$ Yang-Mills theory at $\theta=0$ is trivially gapped with unbroken $CP$, and also that the first-order phase transition does not happen at any $0<\theta<\pi$. 
Let us couple the theory to background $\mathbb{Z}_n$ two-form gauge fields $B$ as we have done in Sec.~\ref{sec:PSU(n)YM}. 

Even after this coupling, $CP$ must be still unbroken by choosing appropriate $p$ at $\theta=0$. If $CP$ is broken after gauging the $\mathbb{Z}_n$ one-form symmetry, then this means that there is a mixed 't Hooft anomaly between the $CP$ symmetry and $\mathbb{Z}_n$ one-form symmetry. Since an 't Hooft anomaly is renormalization group invariant~\cite{Gaiotto:2014kfa, tHooft:1979rat}, there must be a certain degree of freedom carrying the same anomaly and surviving in the infrared limit. 
The assumption on the trivially gapped state claims that there is no such degree of freedom, and thus there must be a way to couple to the $\mathbb{Z}_n$ two-form gauge field $B$ without breaking $CP$. 

Note that the $CP$ transformation flips the sign of the $\int B\wedge B$ term, and effectively $p$ is mapped to $-p$ under the $CP$ transformation. 
Therefore, above discussion claims that we can choose the discrete theta angle satisfying\footnote{The condition derived here is different from and weaker than that given in Ref.~\cite{Gaiotto:2017yup} since we only consider theories on spin manifolds while they consider theories on non-spin manifolds as well as spin ones. It does not affect the consequence about the fate of $CP$ symmetry at $\theta=\pi$. Therefore, discussion given here slightly extends the applicability of the result given in Ref.~\cite{Gaiotto:2017yup} to $SU(n)$ Yang-Mills theories also with adjoint fermions.} 
\be
p=-p\quad \mathrm{mod}\;n
\ee
in order not to break $CP$ at $\theta=0$. Since this has a solution (e.g., $p=0$ mod $n$ is always a solution), the assumption on the gap and unbroken $CP$ at $\theta=0$ is consistent. 
Let us discuss the fate of $CP$ symmetry at $\theta=\pi$. In the $SU(n)$ Yang-Mills theory, $\theta=\pi$ is $CP$-invariant because $CP$ flips $\theta=\pi$ to $\theta=-\pi$ and one can shift $\theta$ to $\theta+2\pi$. After considering the coupling to the $\mathbb{Z}_n$ two-form gauge field $B$, this procedure changes $p$ to $-p-1$ because $CP$ flips $p$ to $-p$ and the $2\pi$ shift of $\theta$ changes $-p$ to $-p-1$ due to the Witten effect. 
In order not to break $CP$ due to the coupling to $B$ at $\theta=\pi$, we must choose the discrete theta angle satisfying 
\be
p=-p-1\quad \mathrm{mod}\;n, 
\ee
but this is inconsistent with our choice at $\theta=0$\footnote{If $n$ is even, $p=-p-1$ (mod $n$) does not have any integer solutions, and thus there is a mixed 't Hooft anomaly. This claims that all the states form pairs under $CP$. For odd $n$, the condition $p=-p-1$ (mod $n$) can be solved by putting $p=(n-1)/2$, and thus there is no 't Hooft anomaly. This means that there exist quasi-vacua that keep $CP$ invariance. However, $p=(n-1)/2$ is not the $CP$-invariant choice at $\theta=0$; the vacuum is not such a $CP$-invariant state at $\theta=\pi$ since we have assumed that there is no first-order phase transition at $0<\theta<\pi$. We shall explain this point in more detail in Sec.~\ref{sec:dual_superconductor}. }. 
Therefore, $CP$ is broken after coupling $SU(n)$ Yang-Mills theory at $\theta=\pi$ to $\mathbb{Z}_n$ background two-form gauge fields  for any choice of the discrete theta angle $p$ preserving $CP$ at $\theta=0$. 

For consistency, there must be some low-energy degrees of freedom in the $SU(n)$ Yang-Mills theory at $\theta=\pi$ that explains the $CP$ breaking after coupling it to $\mathbb{Z}_n$ two-form gauge fields~\cite{Gaiotto:2017yup}. There are several possible candidates for this:
\begin{itemize}
\item The vacua are trivially gapped but degenerate. Each of them breaks $CP$ spontaneously. 
\item The vacuum is gapped with unbroken $CP$ symmetry but described by a nontrivial topological field theory.
\item The theory contains massless excitations.  
\end{itemize}
If one further assumes or proves that the gap does not close at finite $\theta$ and the theory does not show the topological phase transition, $CP$ is broken spontaneously and there is a first-order phase transition at $\theta=\pi$. 
This interesting discussion given in Ref.~\cite{Gaiotto:2017yup} does not rely on any specific microscopic details, and thus the consequence is very general as long as the theory has the $\mathbb{Z}_n$ one-form symmetry (i.e., matters are in the adjoint representation) and satisfies the assumption about the mass gap or topological excitations.

\section{$SU(n)\times SU(n)$ bifundamental gauge theory}\label{sec:SU(n)2_bifundamental}

We consider a gauge theory with the gauge group $SU(n)_1\times SU(n)_2$ and bifundamental matter fields. We use the convention that the gauge fields $a_i$ of $SU(n)_i$ are realized as the traceless and anti-Hermitian $n\times n$ matrix-valued local one-form. 
Our argument in the following is valid for any kinds of the bifundamental matter fields, but, as a specific example,
one can consider single bifundamental Dirac field $\Psi$: $\Psi$ belongs to the fundamental representation of $SU(n)_1$ and to the anti-fundamental representation of $SU(n)_2$, and it is realized as an $n\times n$ matrix-valued four-component Dirac fields. The $SU(n)_1\times SU(n)_2$ gauge transformation $(u_1,u_2)$ acts on $\Psi$ and $a_i$ as $\Psi\mapsto u_1\Psi u_2^{\dagger}$ and $a_i\mapsto u_i a_i u_i^{\dagger}+u_i\diff u_i^{\dagger}$. 
The classical action of the theory is given by
\bea
S&=&-{1\over 2 g_1^2}\int \mathrm{Tr}(G_1\wedge *G_1)-{1\over 2 g_2^2}\int \mathrm{Tr}(G_2\wedge *G_2)+\int\mathrm{Tr}\,\overline{\Psi}(\slashed{D}+m)\Psi\nonumber\\
&&+{\im\theta_1 \over 8\pi^2}\int \mathrm{Tr}(G_1\wedge G_1)+{\im\theta_2\over 8\pi^2}\int\mathrm{Tr}(G_2\wedge G_2), 
\label{eq:SU(n)_bifundamental_action}
\eea
where $G_i$ is the field strength of the $SU(n)_i$ gauge group, 
\be
G_i=\diff a_i+a_i\wedge a_i, 
\ee
and 
\be
\slashed{D}\Psi=\gamma^{\mu}(\p_{\mu}\Psi+a_{1\mu}\Psi-\Psi a_{2\mu}). 
\label{eq:bifund_covariant_derivative}
\ee
We assume that $m>0$, and the matter part does not break $CP$ explicitly. 
We denote the electric and magnetic charge of the $SU(n)_1\times SU(n)_2$ gauge group as $(z_{e1},z_{m1})\oplus (z_{e2},z_{m2})$, then the bifundamental Dirac field has the charge $(1,0)\oplus(n-1,0)$ mod $n$. This theory has fundamental Wilson lines 
\be
W_1(C)=\mathrm{Tr}\left[\mathcal{P}\exp\oint_C a_1\right],\quad  
W_2(C)=\mathrm{Tr}\left[\mathcal{P}\exp\oint_C a_2\right],  
\label{eq:SU(n)_WilsonLine}
\ee
and they have charge $(1,0)\oplus(0,0)$ and $(0,0)\oplus(1,0)$, respectively. $W_1 W_2^{-1}$ has the same charge with the dynamical fermion of this theory. 

Let us describe the ($0$-form) symmetries of this theory. 
$U(1)_V$ is the phase rotation of the fermionic field 
\be
\Psi\mapsto \mathrm{e}^{\im\phi}\Psi, \overline{\Psi}\mapsto \mathrm{e}^{-\im\phi}\overline{\Psi},
\ee 
and this does not act on gauge fields $a_i$. 
If $g_1^2=g_2^2$ and $\theta_1=\theta_2$, there is the $(\mathbb{Z}_2)_I$ symmetry, which interchanges two gauge fields 
\be
a_1\leftrightarrow -a_2^t
\ee
and acts on fermions as $\Psi\mapsto {\Psi}^t$. 
Except for these internal symmetries, there exist usual charge conjugation $C$, parity $P$, and time reversal $T$ symmetries. When the $(\mathbb{Z}_2)_I$ symmetry and charge conjugation is combined, the gauge fields are transformed as $a_1\leftrightarrow a_2$. 

Recall that the $SU(n)$ pure Yang-Mills theory has the electric $\mathbb{Z}_n$ one-form symmetry, and thus this theory has $\mathbb{Z}_n\times \mathbb{Z}_n$ one-form symmetry when the mass $m$ of the Dirac fermion is infinitely large. At finite $m$, the bifundamental Dirac fermion becomes dynamical, and it breaks $\mathbb{Z}_n\times \mathbb{Z}_n$ one-form symmetry to the stabilizer subgroup of $W_1 W_2^{-1}$. The $\mathbb{Z}_n\times \mathbb{Z}_n$ one-form symmetry is explicitly broken to the diagonal $\mathbb{Z}_n$ one-form symmetry. Under this electric one-form symmetry, $W_1$ and $W_2$ have the same charge. 

We study the consistency on the dynamics at $\theta=\pi$ using a mixed 't Hooft anomaly with this electric one-form symmetry, and constrain structures of the phase diagram. For that purpose, we first discuss gauging of $\mathbb{Z}_n$ one-form symmetry. 
For simplicity of discussion, we assume that the vacua are always trivially gapped, and we will study how the first-order phase transition happens as a function of $\theta_1$ and $\theta_2$. 

\subsection{Coupling with the $\mathbb{Z}_n$ two-form gauge fields}

We couple the above $SU(n)_1\times SU(n)_2$ bifundamental gauge theory to $\mathbb{Z}_n$ gauge fields in order to obtain $(SU(n)_1\times SU(n)_2)/(\mathbb{Z}_n)_{\mathrm{diagonal}}$ gauge theory. 
First, we discuss the possible charges of genuine line operators with dynamical matter fields. 
Bifundamental matters have the charge $(1,0)\oplus(-1,0)$, and thus, in order for the line with charge $(z_{e1},z_{m1})\oplus(z_{e2},z_{m2})$ to be a genuine line, the Dirac quantization condition requires 
\be
{1\over n}(z_{m1}-z_{m2})=0\quad \mathrm{mod}\; 1. 
\ee
This means that two magnetic charges modulo $n$ for $SU(n)_{1,2}$ gauge groups must be the same. 
Let us consider the case where the theory have a genuine line operator with the magnetic charge $z_{m1}=z_{m2}=1$. 
The Dirac quantization further restricts the possible purely electric genuine lines.  To see it, let $(z_{e1},0)\oplus(z_{e2},0)$ be a charge of the genuine line, then we get 
\be
z_{e1}+z_{e2}=0 \quad \mathrm{mod}\; n,
\ee
from the Dirac quantization. As a result, the purely electric lines are given by $(W_1W_2^{-1})^{k}$ for $k=0,1,\ldots, n-1$. 

We shall obtain such a theory by coupling the $SU(n)\times SU(n)$ bifundamental gauge theory to a $\mathbb{Z}_n$ topological field theory. 
We introduce the $\mathbb{Z}_n$ two-form gauge field $B$, and its classical action is given by the same action in (\ref{eq:ZnTFT}): 
\be
S_{\mathrm{TFT}}={\im\over 2\pi}\int F\wedge (\diff A+n B)+{\im np\over 4\pi }\int B\wedge B. 
\label{eq:ZnTFT_re}
\ee
Here, $A$ and $B$ are $U(1)$ one-form and two-form gauge fields, and the equation of motion for $F$ requires $nB=-\diff A$, which makes $B$ a $\mathbb{Z}_n$ two-form gauge field. 
We consider theories only on spin manifolds since we would like to include the case where bifundamental matters are Dirac fermions, then the parameter $p$ must be an integer mod $n$: 
The condition on $p$ being an integer comes from the requirement on the $U(1)$ one-form gauge invariance of (\ref{eq:ZnTFT_re}). $p$ is identified with $p+n$ since integration out of $F$ yields
\be
S_{\mathrm{TFT}}=2\pi \im {p\over n}\left({1\over 2}\int{\diff A\over 2\pi}\wedge {\diff A\over 2\pi}\right),
\ee 
and difference of $p$ by multiples of $n$ gives the difference of $S_{\mathrm{TFT}}$ in $2\pi\im \mathbb{Z}$. Hence, it does not affect the result in quantum theories. 

To couple $SU(n)$ gauge fields $a_1$, $a_2$ to $B$, we first extend the gauge group $SU(n)_1\times SU(n)_2$ to 
\be
{SU(n)_1\times SU(n)_2\times U(1)\over \mathbb{Z}_n}, 
\ee
and replace the $SU(n)$ gauge fields $a_1$ and $a_2$ by $U(n)$ gauge fields 
\be
\calA_1=a_1+{1\over n}A \bm{1}_n, \, \calA_2=a_2+{1\over n}A \bm{1}_n. 
\ee
The $U(1)$ gauge field $A=\mathrm{Tr}(\calA_1)=\mathrm{Tr}(\calA_2)$ is the same with the one that appears in (\ref{eq:ZnTFT_re}), and this creates the coupling of theories that we want. 
This $U(1)$ gauge field $A$ does not couple to bifundamental fields, and it can be easily checked by an explicit form of the covariant derivative (\ref{eq:bifund_covariant_derivative}). 

We construct the Wilson and 't Hooft line operators, which need not be genuine but must be gauge-invariant. After that, we study the spectrum of genuine line operators to check whether we have obtained the $(SU(n)_1\times SU(n)_2)/\mathbb{Z}_n$ gauge theory. The former definitions of Wilson lines in (\ref{eq:SU(n)_WilsonLine}) are no longer gauge-invariant after gauging the $\mathbb{Z}_n$ one-form symmetry. 
Let $\Sigma$ be a two-dimensional surface with $C=\p \Sigma$, and the gauge-invariant Wilson loops are defined by
\bea
W_1(C,\Sigma)&=&\mathrm{Tr}\left[\calP \exp\left(\oint_C a_1\right)\right]\exp\left(\oint_C{1\over n}A+\int_\Sigma B\right),\\ 
W_2(C,\Sigma)&=&\mathrm{Tr}\left[\calP \exp\left( \oint_C a_2\right)\right]\exp\left(\oint_C{1\over n}A+\int_\Sigma B\right). 
\eea
The magnetic one with charge $(0,1)\oplus (0,1)$ is also defined by 
\be
H(C,\Sigma)=\exp\left( \int_{\Sigma}(F+p B)\right). 
\ee
Using Wilson lines, the genuine line operator of charge $(1,0)\oplus (-1,0)$ is given by 
\be
W_1(C,\Sigma)W_2(C,\Sigma)^{-1}=\mathrm{Tr}\left[\calP \exp\left(\oint_C a_1\right)\right]\left(\mathrm{Tr}\left[\calP \exp\left( \oint_C a_2\right)\right]\right)^{-1}. 
\ee
We can also construct a dyonic genuine line object, 
\be
H(C,\Sigma)W_1(C,\Sigma)^{-p}=\exp\left(\int_\Sigma F\right) \left(\mathrm{Tr}\left[\mathcal{P}\exp\oint_C\left( a_1+{1\over n}A\right)\right]\right)^{-p}, 
\ee
which has the charge $(-p,1)\oplus(0,1)$. By multiplying $(W_1 W_2^{-1})^k$ to it, we can generally obtain the genuine line operator $H W_1^{-p}(W_1 W_2^{-1})^k$ with the charge $(-p+k,1)\oplus(-k,1)$ mod $n$. The discrete theta angle $p$ designates the sum of electric charge for the genuine dyonic particles with the magnetic charges $1$. 

Since the topological $\theta$ angle is the central issue of our discussion, we compute how it is changed after the gauging in an explicit manner. 
In oder to maintain the $1$-form gauge invariance, we should replace the gauge field strength $G_{1}$ and $G_2$ by $\calG_{1}+B$ and $\calG_{2}+B$, respectively, where $\calG_{i}$ are the $U(n)$ field strengths of $\calA_i$; $\calG_i=\diff \calA_i+\calA_i\wedge\calA_i$. 
As a result, the topological $\theta$ term becomes 
\bea
S_{\theta}&=&\sum_{i=1,2}{\im \theta_i\over 8\pi^2}\int \mathrm{Tr}\left[(\calG_i+B)\wedge (\calG_{i}+ B)\right]\nonumber\\
&=&\sum_{i=1,2}{\im \theta_i\over 8\pi^2}\int\left\{\mathrm{Tr}(\calG_i\wedge \calG_i)+2B\wedge \mathrm{Tr}(\calG_i)+ n B\wedge B\right\}. 
\eea
Using the equation of motion of $F$, $\mathrm{Tr}(\calG_i)=\diff A=-n B$, we obtain
\be
S_{\theta}=\sum_{i=1,2}{\im \theta_i\over 8\pi^2}\int \left\{\mathrm{Tr}(\calG_i\wedge \calG_i)-n B\wedge B\right\}. 
\label{eq:SU(n)_bifund_topological}
\ee

Using the consistency of the local counter term $p$ with the $CP$ symmetry, we will discuss the possible phase structure of the $SU(n)\times SU(n)$ bifundamental gauge theories in the following sections. 

\subsection{Spontaneous $CP$ breaking at $(\theta_1,\theta_2)=(\pi,0)$ and $(0,\pi)$}

We first constrain the possible dynamics of bifundamental gauge theories at $(\theta_1,\theta_2)=(\pi,0)$ or $(\theta_1,\theta_2)=(0,\pi)$. We follow the same logic given in Ref.~\cite{Gaiotto:2017yup}, and start with the assumption that the vacuum at $\theta_1=\theta_2=0$ is trivially gapped without breaking the $CP$ symmetry. 
Therefore, there must be a way to gauge other symmetries without breaking the $CP$ symmetry at $\theta$=0 by using an 't Hooft anomaly matching condition. 
Especially when gauging the electric $\mathbb{Z}_n$ symmetry, the local counter term ${\im n p\over 4\pi}\int B\wedge B$ can be chosen to be $CP$ invariant from this argument, and such $p$ must satisfy
\be
2p=0\quad \mathrm{mod}\; n, 
\ee
since $\int B\wedge B$ flips its sign under the $CP$ transformation.

We further assume that the vacua are always trivially gapped and that there is a way to continuously connect $(\theta_1,\theta_2)=(0,0)$ and $(\theta_1,\theta_2)=(\pi,0),\,(0,\pi)$ without phase transitions. 
We will show that there exists first-order phase transition associated with the spontaneous $CP$ breaking at $(\theta_1,\theta_2)=(\pi,0)$ and at $(\theta_1,\theta_2)=(0,\pi)$ under this assumption.

Let us discuss the $CP$ symmetry at $\theta_1=\pi$ with $\theta_2=0$ after gauging the $\mathbb{Z}_n$ one-form symmetry. 
Since $CP$ flips the orientation, $p$ and $\theta_i$ change their signs and become $-p$ and $-\theta_i$, respectively. In oder to consider the theory at $\theta_1=\pi$, we must consider not only the change $\theta_1=\pi\mapsto \theta_1'=-\pi$ but also the shift $\theta_1'=-\pi\mapsto \theta_1'+2\pi =\pi$ to discuss its $CP$ invariance. 
Under these transformations, the topological $\theta$ term is changed by 
\be
\Delta S_{\theta}={2\pi \im \over 8\pi^2}\int \mathrm{Tr}(\calG_1\wedge \calG_1)-{\im n\over 4\pi}\int B\wedge B. 
\ee
The first term is in $2\pi \im \mathbb{Z}$, and thus does not affect the path integral. The second term shifts the value of $p$ by $-1$. 
As a result, $p$ is changed to $p\mapsto -p-1$ under the $CP$ transformation at $(\theta_1,\theta_2)=(\pi,0)$, and thus the condition for the $CP$ invariance at $(\theta_1,\theta_2)=(\pi,0)$ after gauging is given by 
\be
p=-p-1\quad \mathrm{mod}\; n. 
\ee
For even $n$, there is no such integer $p$. Therefore, there is an 't Hooft anomaly, and all the quasi-vacua must form pairs under $CP$ or become gapless to saturate the anomaly. 
For odd $n$, $p=(n-1)/2$ satisfies this condition, but it is inconsistent with the choice of $p$ at $\theta_1=\theta_2=0$.  
Since we put an assumption that a vacuum at $(\theta_1,\theta_2)=(\pi,0)$ is continuously connected to the $CP$-invariant vacuum at $\theta_1=\theta_2=0$, consistency condition requires the existence of low-energy degrees of freedom to saturate this inconsistency, such as degenerate vacua or massless excitations. 
Since we have also assumed that the mass gap does not close, there exists the first-order phase transition at $(\theta_1,\theta_2)=(\pi,0)$ in both cases associated with the spontaneous $CP$ breaking. 

The same argument holds for $(\theta_1,\theta_2)=(0,\pi)$, and we can argue the spontaneous $CP$ breaking there. 

\subsection{Vacuum structure around $\theta_1=\theta_2=\pi$}

\begin{figure}[t]
\centering
\includegraphics[scale=0.7]{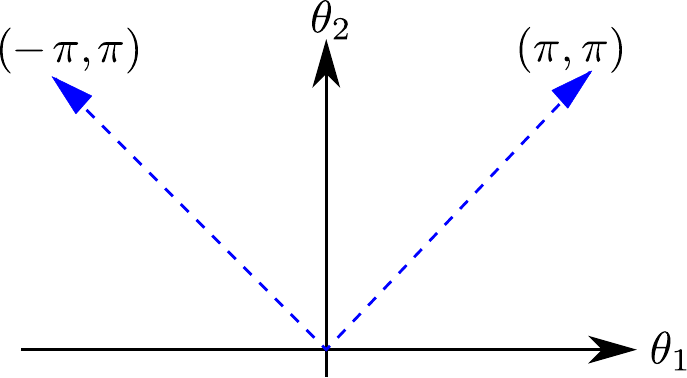}
\caption{Two different paths connecting $(\theta_1,\theta_2)=(0,0)$ and $(\theta_1,\theta_2)=(\pi,\pi)\sim (-\pi,\pi)$. }
\label{fig:bifund_theta_path}
\end{figure}

Let us next discuss the consistency condition for $CP$ at $\theta_1=\theta_2=\pi$. 
This case is somewhat tricky, since there are two topologically distinct ways that connect $(\theta_1,\theta_2)=(\pi,\pi)$ and $(\theta_1,\theta_2)=(0,0)$ (See Fig.~\ref{fig:bifund_theta_path}). 
Since both $\theta_1$ and $\theta_2$ are $2\pi$ periodic for the gauge group $SU(n)\times SU(n)$, $\theta_1=\theta_2=\pi$ and $\theta_1=-\theta_2=-\pi$ are equivalent. 
We will discuss whether we encounter the first-order phase transition when changing $\theta_1$ and $\theta_2$ continuously from $(\theta_1,\theta_2)=(0,0)$ to $(\theta_1,\theta_2)=(\pi,\pi)$ or $(-\pi,\pi)$. 

Let us consider the case $(\theta_1,\theta_2)=(\pi,\pi)$. After gauging the $\mathbb{Z}_n$ symmetry, we must use the $2\pi$ periodicity of both $\theta_1$ and $\theta_2$ to discuss the $CP$ symmetry, and under these shifts the topological term (\ref{eq:SU(n)_bifund_topological}) is changed by  
\be
\Delta S_{\theta}={2\pi \im \over 8\pi^2}\sum_{i=1,2}\int \mathrm{Tr}(\calG_i\wedge \calG_i)-{2\im n\over 4\pi}\int B\wedge B. 
\ee
On spin manifolds, the first term is in $2\pi \im \mathbb{Z}$ and does not affect the path integral. 
It thus changes $p$ to $p-2$. One can understand this from the spectrum of genuine line operators. Originally, spectrum of genuine line operators are given by $(-p+k,1)\oplus(-k,1)$ mod $n$ with $k=0,\ldots,n-1$. Since they have the monopole charge $1$, the $2\pi$ shift of $\theta_{1,2}$ causes the shift of charge $(-p+k+1,1)\oplus (-k+1,1)$ due to the Witten effect, and they become $(-p+2+k',1)\oplus(-k',1)$ mod $n$ with $k'=0,\ldots,n-1$ by putting $k'=k-1$. 
Notice that the spectrum is not changed only when $n=2$, and this will become important  for our result.

Let us consider whether there is a way to gauge the electric one-form symmetry without breaking the $CP$ invariance at $\theta_1=\theta_2=\pi$. After gauging, we have a local counter term ${\im np\over 4\pi}\int B\wedge B$, which flips the sign under $CP$. It can be described effectively by the map $p\mapsto -p$, and $\theta_{1,2}=\pi\mapsto -\pi$. To get the original topological angle, we perform the $2\pi$ shift of both $\theta_1$ and $\theta_2$ that changes $-p\mapsto -p-2$. 
As a result, the $CP$ invariance at $\theta_1=\theta_2=\pi$ after gauging requires to choose $p$ satisfying
\be
p=-p-2\quad \mathrm{mod}\; n.
\ee
This always has the integer solution, and thus there is an $CP$-invariant quasi-vacuum which may or may not be the true vacuum. 
Let us next discuss the global consistency condition. 
If $\theta_1=\theta_2=0$ and $\theta_1=\theta_2=\pi$ can be continuously connected without breaking $CP$ at $\theta_1=\theta_2=\pi$, then the integer solution $p$ at $\theta_1=\theta_2=\pi$ must also be consistent with the $CP$-invariant regularization at $\theta_1=\theta_2=0$; this says that 
\be
2=0\quad \mathrm{mod}\; n. 
\label{eq:condition_CP_unbroken}
\ee
The vacuum at $\theta_1=\theta_2=0$ can be continuously changed to the $CP$-invariant vacuum at $\theta_1=\theta_2=\pi$ without closing the mass gap only if this condition holds.

For $n\ge 3$, the above global consistency relation cannot be true. One possibility is that the vacua at $\theta_{1,2}=0$ and $\theta_{1,2}=\pi$ are separated  by first-order phase transitions. Another possibility is that the vacuum at $\theta_{1,2}=\pi$ breaks $CP$ spontaneously to saturate the inconsistency. 
For $n=2$, we cannot impose any constraints on the state at $\theta_1=\theta_2=\pi$ from our argument, and basically any possibilities are allowed\footnote{This might be because we consider theories defined only on spin manifolds. If we restrict our attention to theories without fermions, then theories can be defined also on non-spin manifolds. We can repeat the same argument for non-spin cases at least formally just by changing the identification of the discrete theta angle from $p\sim p+n$ to $p\sim p+2n$. The necessary condition for unbroken $CP$ given by (\ref{eq:condition_CP_unbroken}) becomes $2=0$ mod $2n$, and then we would find that $CP$ must be broken for all $n\ge 2$. 
Since we are not familiar with non-spin case, however, let us leave it as a speculative remark. }. 

%
%
%
%

\begin{figure}[t]
\centering
\begin{minipage}{.45\textwidth}
\subfloat[$CP$ is unbroken at $(\theta_1,\theta_2)=(\pi,\pi)$]{
\includegraphics[scale=0.5]{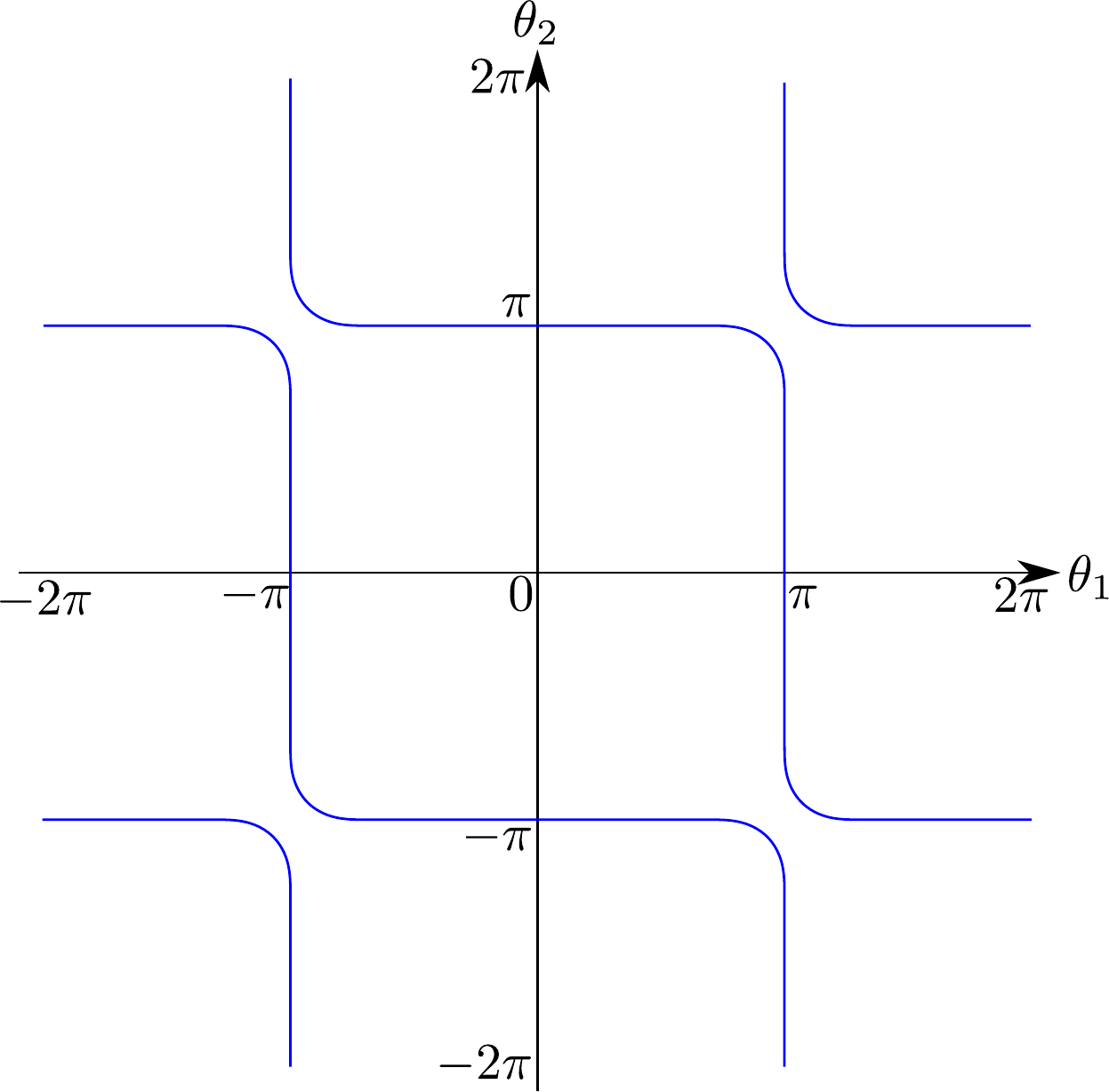}
\label{fig:bifund_phase_boundary1}
}\end{minipage}\quad
\begin{minipage}{.45\textwidth}
\subfloat[$CP$ is broken at $(\theta_1,\theta_2)=(\pi,\pi)$]{
\includegraphics[scale=0.5]{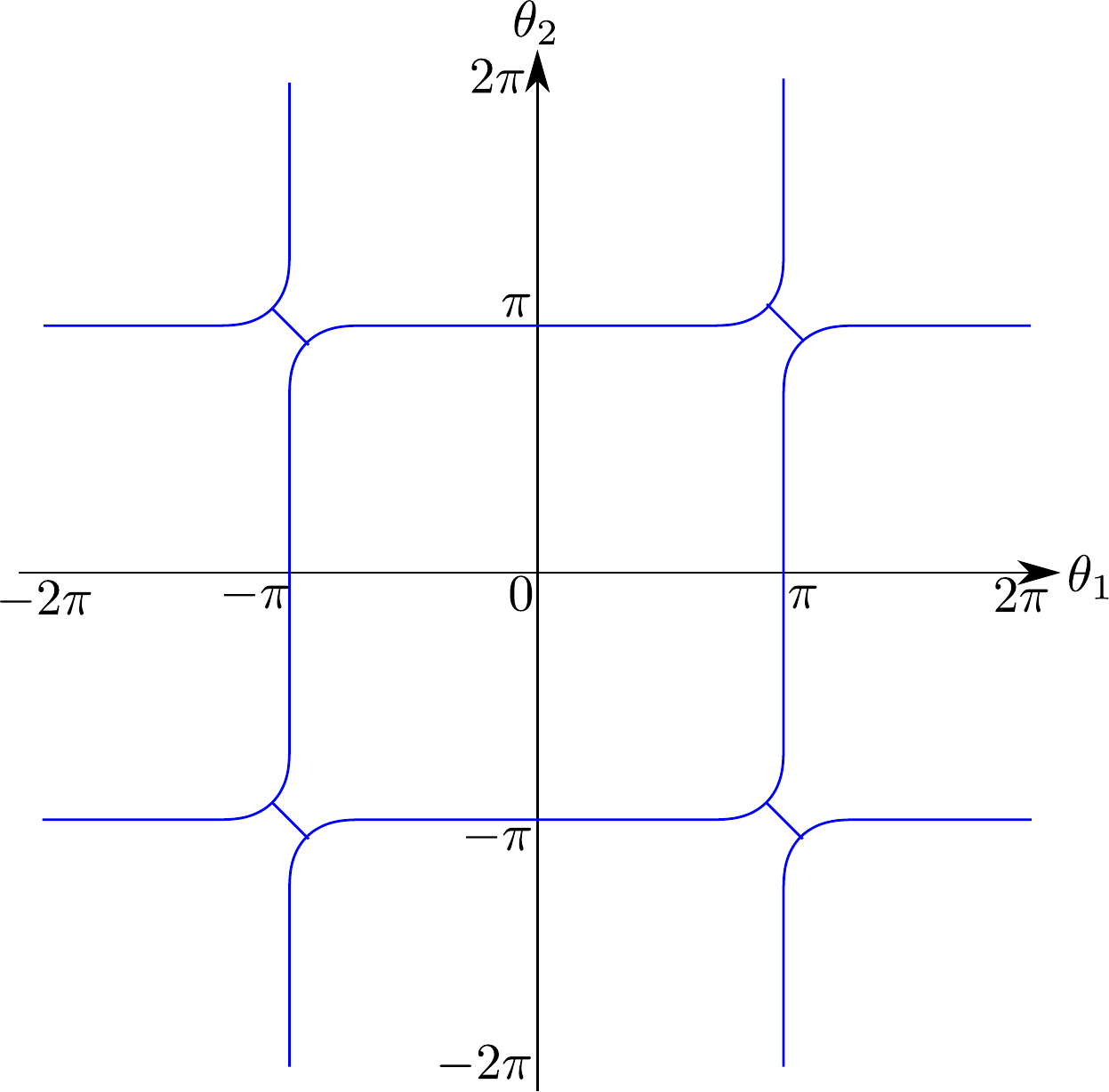}
\label{fig:bifund_phase_boundary2}
}\end{minipage}
\caption{Possible phase boundaries of $SU(n)\times SU(n)$ bifundamental gauge theories in the $\theta_1$-$\theta_2$ plane ($n\ge 3$). }
\label{fig:bifund_phase_boundary}
\end{figure}

Next, let us consider what happens when connecting $\theta_1=\theta_2=0$ and $\theta_1=-\theta_2=-\pi$. 
In this case, the $CP$ transformation at $(\theta_1,\theta_2)=(-\pi,\pi)$ is associated with the shift $\theta_1\mapsto \theta_1-2\pi$ and $\theta_2\mapsto \theta_2+2\pi$. The change of the topological term under these shifts is given by 
\be
\Delta S_{\theta}={\im\over 4\pi}\int \mathrm{Tr}(-\calG_1\wedge \calG_1+\calG_2\wedge \calG_2)\in 2\pi\im\mathbb{Z}. 
\ee 
Therefore it does not affect the path integral at all. 
In this case, the $CP$ transformation changes $p\mapsto -p$ mod $n$ as in the case of $\theta_1=\theta_2=0$. 
When connecting $\theta_1=\theta_2=0$ and $\theta_1=-\theta_2=-\pi$, the global consistency holds and thus the vacua can be continuously connected without the phase transition and $CP$ needs not be broken at $\theta_1=-\theta_2=-\pi$. 

By combining the result and respecting the $2\pi$ periodicity of $\theta_{1,2}$, we obtain Fig.~\ref{fig:bifund_phase_boundary} as a possible phase boundary of the first-order phase transition in the $\theta_1$-$\theta_2$ plane when $n\ge 3$. Whether the phase boundary opens at $(\theta_1,\theta_2)=(\pi,\pi)$ depends on details of the dynamics such as matter contents. 
In Fig.~\ref{fig:bifund_phase_boundary1}, we consider the possibility when the $CP$ symmetry is unbroken at $\theta_1=\theta_2=\pi$. 
In this case, the first-order phase transition line must separate $\theta_1=\theta_2=0$ and $\theta_1=\theta_2=\pi$, but $\theta_1=\theta_2=0$ and $\theta_1=-\theta_2=\pi$ can be smoothly connected. 
In Fig.~\ref{fig:bifund_phase_boundary2}, the $CP$ symmetry is spontaneously broken at $\theta_1=\theta_2=\pi$, and thus there is a first-order phase transition line around it. 
In this case, $\theta_1=\theta_2=0$ and $\theta_1=-\theta_2=\pi$ would be separated by another first-order line because the vacuum at $\theta_1=\theta_2=0$ is continuously connected to the $CP$-invariant quasi-vacuum at $\theta_1=-\theta_2=\pi$ but not to the true $CP$-broken vacuum according to the global consistency relation. 

One may wonder whether the first-order phase transition line in Fig.~\ref{fig:bifund_phase_boundary} can terminate so that one can smoothly change $(\theta_1,\theta_2)$ from $(0,0)$ to $(2\pi,0)$ without phase transitions. In Fig.~\ref{fig:bifund_phase_boundary}, this is impossible and we claim that it is a general result for $n\ge 3$. 
By repeating the same argument on the $CP$ transformation after gauging the $\mathbb{Z}_n$ symmetry, the condition for the $CP$ invariance at $(\theta_1,\theta_2)=(2\pi,0)$ is given by $p=-p-2$ mod $n$. 
Of course this has the solution, but it is inconsistent with the $CP$ invariant choice at $(\theta_1,\theta_2)=(0,0)$ when $n\ge 3$. 
This means that if we could connect $(\theta_1,\theta_2)=(0,0)$ and $(2\pi,0)$ without any phase transition and without closing the mass gap, then $CP$ must be spontaneously broken at $(\theta_1,\theta_2)=(2\pi,0)$ but this is the contradiction because $(\theta_1,\theta_2)=(0,0)$ and $(2\pi,0)$ must be equivalent for $SU(n)\times SU(n)$ gauge theories. 
If we further assume that the mass gap does not close at generic $(\theta_1,\theta_2)$, then $(\theta_1,\theta_2)=(0,0)$ and $(2\pi,0)$ must be separated by the first-order phase transition line when $n\ge 3$. 
For $n=2$, this is not the case.

\subsection{Interpretation via the dual superconductor picture}\label{sec:dual_superconductor}
The purpose of this section is to understand the result intuitively from the dual superconductor model of confinement~\cite{Nambu:1974zg, tHooft:1981bkw, Mandelstam:1974pi}. 
Let us first consider the case $m\to \infty$ and bifundamental matters decouple. Then, we have two decoupled $SU(n)$ Yang-Mills theories, so let us start with the discussion for the $SU(n)$ Yang-Mills theory. 

\subsubsection{$SU(n)$ Yang-Mills theory}

Following the dual superconductor model, we assume that confinement of $SU(n)$ Yang-Mills theory on $\mathbb{R}^4$ is caused by condensation of magnetic monopoles or dyons. Let us say that their charges are given by $(-k,1)$ mod $n$ with $k=0,1,\ldots,n-1$ at $\theta=0$.  This assumes that all the Wilson loops with nontrivial center elements obey the area law. 
There are $n$ candidates of condensed particles, and correspondingly there are $n$ different quasi-vacua. 
To be specific, let us assume that the magnetic monopole with charge $(0,1)$ condenses at $\theta=0$ in the true vacuum. 
Now, we turn on the finite topological $\theta$ angle, and the Witten effect shifts charges of dyons to $(-k+\theta/2\pi,1)$. Since the charge of each dyon goes back to its original value only after the shift of $2\pi n$, each branch of quasi-vacua are $2\pi n$ periodic in $\theta$ instead of $2\pi$ periodic. 
However, the true vacuum must be $2\pi$ periodic in terms of $\theta$, so there must be some jump among quasi-vacua between $0<\theta<2\pi$. 

\begin{figure}[t]
\centering
\includegraphics[scale=.4]{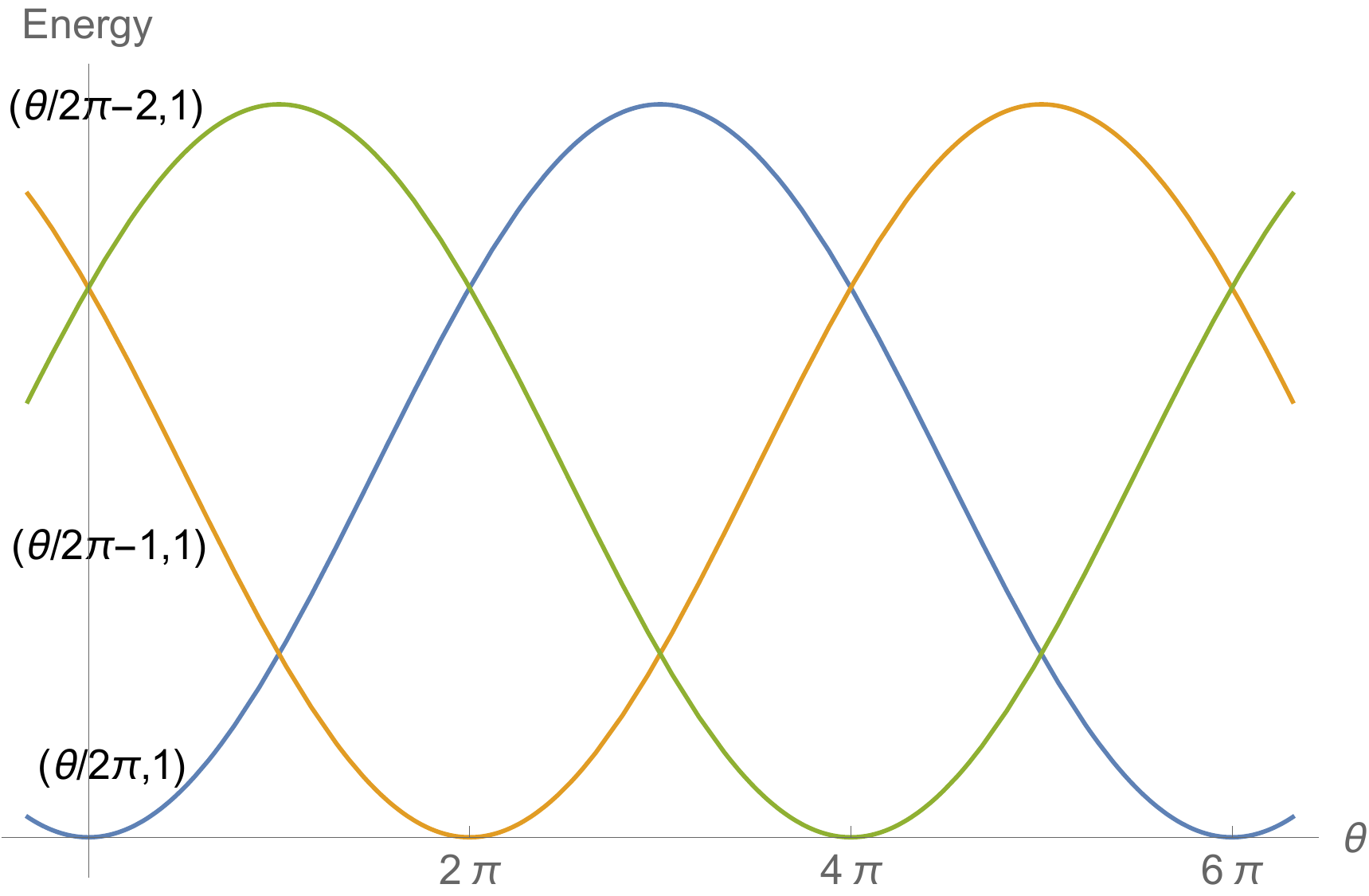}
\caption{Schematic figure on the ground state energy $E(\theta)$ of the $SU(n)$ Yang-Mills theory based on the dual superconductor model for $n=3$. There are $n$ different branches labeled by the condensed charge $(\theta/2\pi -k,1)$ of dyons ($k=0,1,\ldots,n-1$), and each branch is $2\pi n$ periodic. }
\label{fig:theta_SU(n)_YM}
\end{figure}

Let us pay attention to the charge at $\theta=\pi$. Assuming that no phase transition occurs for $0<\theta<\pi$, then the charge of condensed particles (magnetic monopole at $\theta=0$) becomes $(\theta/2\pi,1)$ due to the Witten effect. 
It is not invariant under the $CP$ transformation at $\theta=\pi$ although the theory is $CP$ invariant. Under the $CP$ transformation, the charge $(1/2,1)$ is mapped to $(-1/2,1)$, and thus the quasi-vacua with charges $(\pm 1/2,1)$ must have the same energy because of the $CP$ symmetry of the theory. 
Therefore, the first-order phase transition occurs at $\theta=\pi$, and the true vacua jumps from the branch with the condensed charge $(\theta/2\pi,1)$ to the another branch with the condensed charge $(-1+\theta/2\pi,1)$ (see Fig.~\ref{fig:theta_SU(n)_YM}). 
This is how the $CP$ symmetry is spontaneously broken at $\theta=\pi$ for pure $SU(n)$ Yang-Mills theory in the dual superconductor scenario. 

To summarize the case for the $SU(n)$ Yang-Mills theory, let us denote $E_0(\theta)$ as the energy of the quasi-vacuum with the condensed charge $(\theta/2\pi,1)$. $CP$ symmetry tells us that $E_0(\theta)=E_0(-\theta)$, and $n$-ality shows that $E_0(\theta+2\pi n)=E_0(\theta)$. There are $n$ candidates for the condensate, $(-k+\theta/2\pi,1)$ with $k=0,\ldots,n-1$, and the energy of the true vacuum is 
\be
E_{SU(n)}(\theta)=\mathrm{min}\{E_0(\theta-2\pi k) \,|\, k=0,1,\ldots,n-1\}. 
\ee 
If $E_0$ is smooth, it is natural to have the bump for $E(\theta)$ at $\theta=\pi$, at which the branch jumps from $E_0(\theta)$ to $E_0(\theta-2\pi)$ with the first-order phase transition (see Fig.~\ref{fig:theta_SU(n)_YM}). 
In the large-$n$ limit, it is well established that the Yang-Mills vacuum is described by the minimum of $n$ branches~\cite{Witten:1980sp, Witten:1998uka}. 

Before going to the case of bifundamental gauge theories, let us deepen our understandings on the meaning of the global consistency condition about 't Hooft anomaly matching. 
$CP$ invariance at $\theta=\pi$ requires that $p=-p-1$ mod $n$, and it cannot be solved for even $n$. 
In the dual superconductor picture, the condensed particles at $\theta=\pi$ have charges $(\pm1/2,1),\ldots,(\pm(n-1)/2,1)$ and they form $n/2$ $CP$-invariant pairs. 
Including quasi-vacua, no states can be invariant under $CP$, and this is suggested by the 't Hooft anomaly. 
Next, let us consider the case of odd $n$, then 't Hooft anomaly does not exist by setting $p=(n-1)/2$. 
The condensed charges are given by $(\pm1/2,1),\ldots,(\pm (n-2)/2,1)$ and $(n/2,1)$. 
Since the quasi-vacuum with the condensed charge $(n/2,1)$ is invariant under $CP$ (see Fig.~\ref{fig:theta_SU(n)_YM}), one cannot argue the spontaneous $CP$ breaking at $\theta=\pi$ without putting another assumption. 
The point is that the state with the charge $(n/2,1)$ at $\theta=\pi$ is not continuously connected to the vacuum with the charge $(0,1)$ at $\theta=0$, so the absence of the first-order phase transition at $0<\theta<\pi$ can purge this state from our consideration on vacua. 
In the language of the consistency condition, this is implied by the fact that there is no common integer $p$ for the $CP$ invariance at $\theta=0$ and $\theta=\pi$. 

\subsubsection{$SU(n)\times SU(n)$ bifundamental gauge theories}

Let us now discuss $SU(n)\times SU(n)$ Yang-Mills theory. Considering the limit $m\to \infty$ so that bifundamental matters decouple, we just have two copies of the above argument. 

\begin{figure}[t]
\centering
\includegraphics[scale=.55]{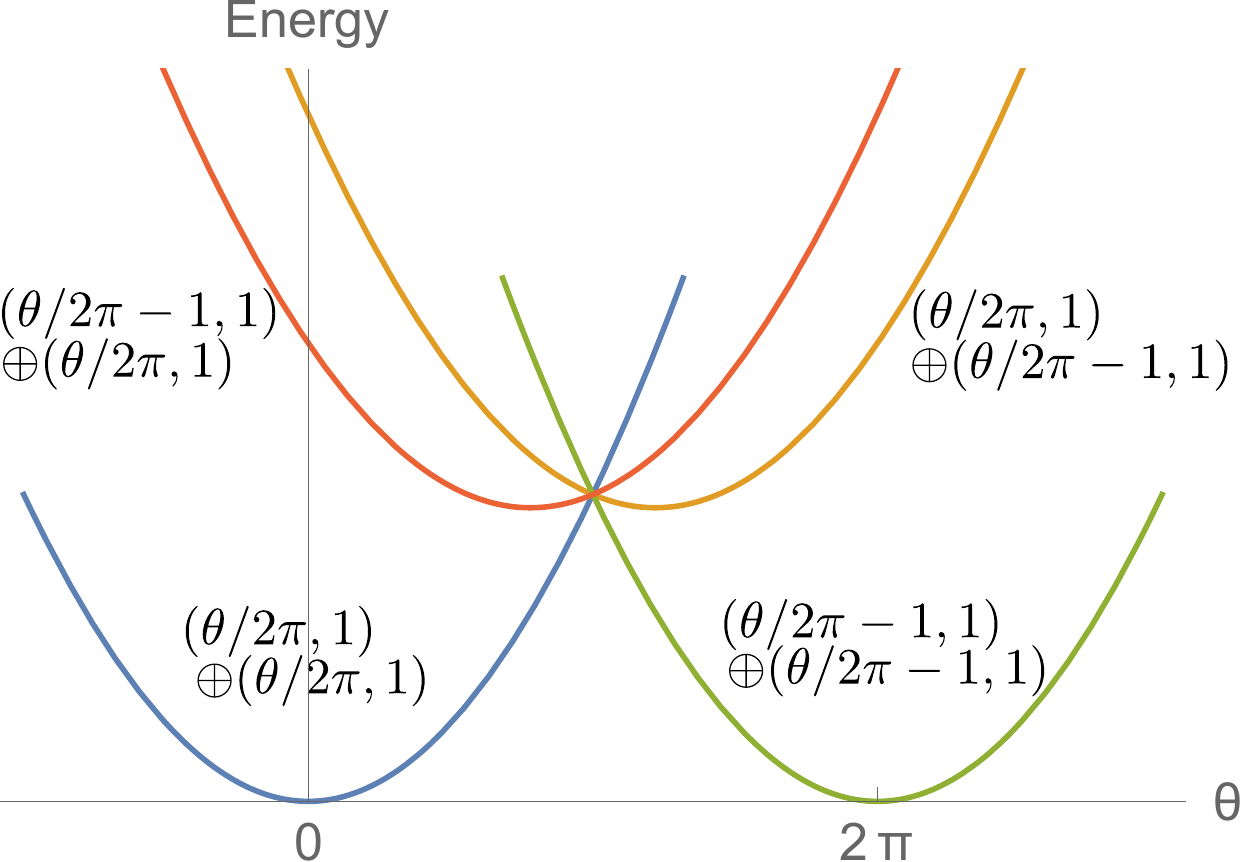}
\caption{Energies of the quasi-vacua of the $SU(n)\times SU(n)$ gauge theory in the limit of $m\to \infty$ when $\theta=\theta_1=\theta_2$. }
\label{fig:Bifund_theta01}
\end{figure}

We first connect $\theta_1=\theta_2=0$ and $\theta_1=\theta_2=\pi$. We select the path $\theta_1=\theta_2$ for instance and denote the common angle as $\theta:=\theta_1=\theta_2$. 
We now have $n^2$ candidates for the condensed particles with the charge $(-k+\theta/2\pi,1)\oplus (-\ell+\theta/2\pi,1)$ mod $n$ with $k,\ell=0,1,\ldots,n-1$, and thus the ground-state energy is given by 
\be
E_{SU(n)\times SU(n)}(\theta)=\mathrm{min}\{E_0(\theta-2\pi k)+E_0(\theta-2\pi \ell)\,|\,k,\ell=0,\ldots,n-1\}. 
\ee 
By assumption that the monopole $(0,1)$ condenses for the $SU(n)$ Yang-Mills theory at $\theta=0$, the quasi-vacuum with $(\theta/2\pi,1)\oplus (\theta/2\pi,1)$ is selected when $\theta$ is close to zero. 
At $\theta=\pi$, $CP$ is broken and the ground state must be at least two-fold degenerate. In our limit $m\to \infty$, there is four-fold degeneracy at $\theta=\pi$, and the condensed charges for those four states are 
\bea
&&(\theta/2\pi,1)\oplus (\theta/2\pi,1), \; (\theta/2\pi-1,1)\oplus (\theta/2\pi-1,1), \nonumber\\
&&(\theta/2\pi-1,1)\oplus (\theta/2\pi,1), \;(\theta/2\pi,1)\oplus (\theta/2\pi-1,1). 
\eea
This is because the $CP$ symmetry is extended to $\mathbb{Z}_2\times \mathbb{Z}_2$ from $\mathbb{Z}_2$ in the limit $m\to \infty$ as a result of the decoupling between two $SU(n)$ Yang-Mills theories. 
If we assume that $E_0(\theta)$ is smooth and monotonically increasing for $0<\theta<2\pi$, there is the first-order phase transition from the sate with condensed charge $(\theta/2\pi,1)\oplus (\theta/2\pi,1)$ to the another one with $(\theta/2\pi-1,1)\oplus (\theta/2\pi-1,1)$ at $\theta=\pi$ (see Fig.~\ref{fig:Bifund_theta01}). 

Let us turn on finite $m$ and make the bifundamental matters dynamical. Then, the $CP$ symmetry becomes $\mathbb{Z}_2$ and the accidental four-fold degeneracy at $\theta=\pi$ must be resolved. Let us first notice that states with the charge $(\theta/2\pi,1)\oplus (\theta/2\pi,1)$ and $(\theta/2\pi-1,1)\oplus (\theta/2\pi-1,1)$ cannot be mixed by dynamical bifundamental fields since the difference of their charges is different from the bifundamental charge $(1,0)\oplus (-1,0)$. 
On the other hand, the difference of two charges $(\theta/2\pi,1)\oplus (\theta/2\pi-1,1)$ and $(\theta/2\pi-1,1)\oplus (\theta/2\pi,1)$ is given by $(1,0)\oplus(-1,0)$, and this is nothing but the charge of dynamical matter fields $\Psi$. 
These states can be mixed as a result of interacting bifundamental matters, which leads to the non-degenerate quasi-vacuum with the mass gap. 

\begin{figure}[t]\centering
\begin{minipage}{.45\textwidth}
\subfloat[$CP$ is unbroken at $\theta=\pi$]{
\includegraphics[scale=0.5]{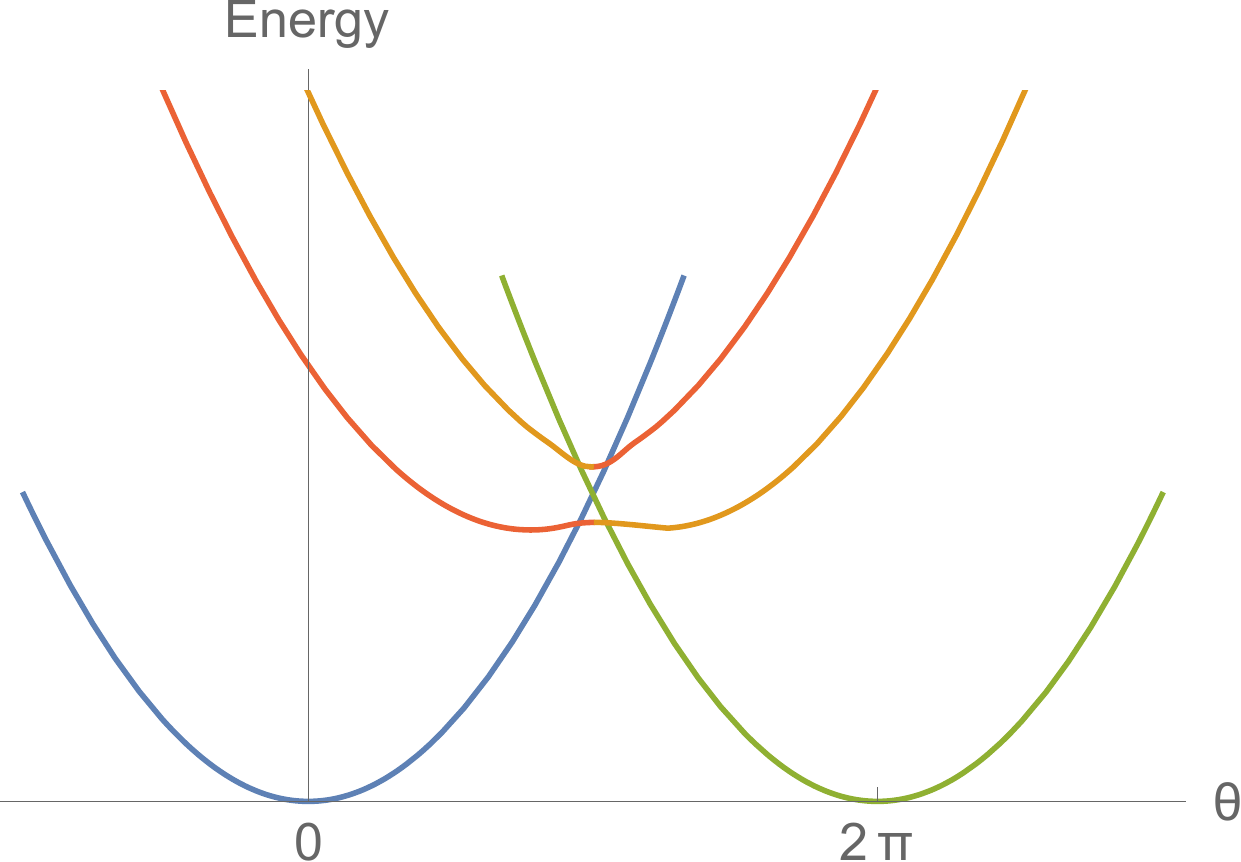}
\label{fig:Bifund_theta02a}
}\end{minipage}\quad
\begin{minipage}{.45\textwidth}
\subfloat[$CP$ is broken at $\theta=\pi$]{
\includegraphics[scale=0.5]{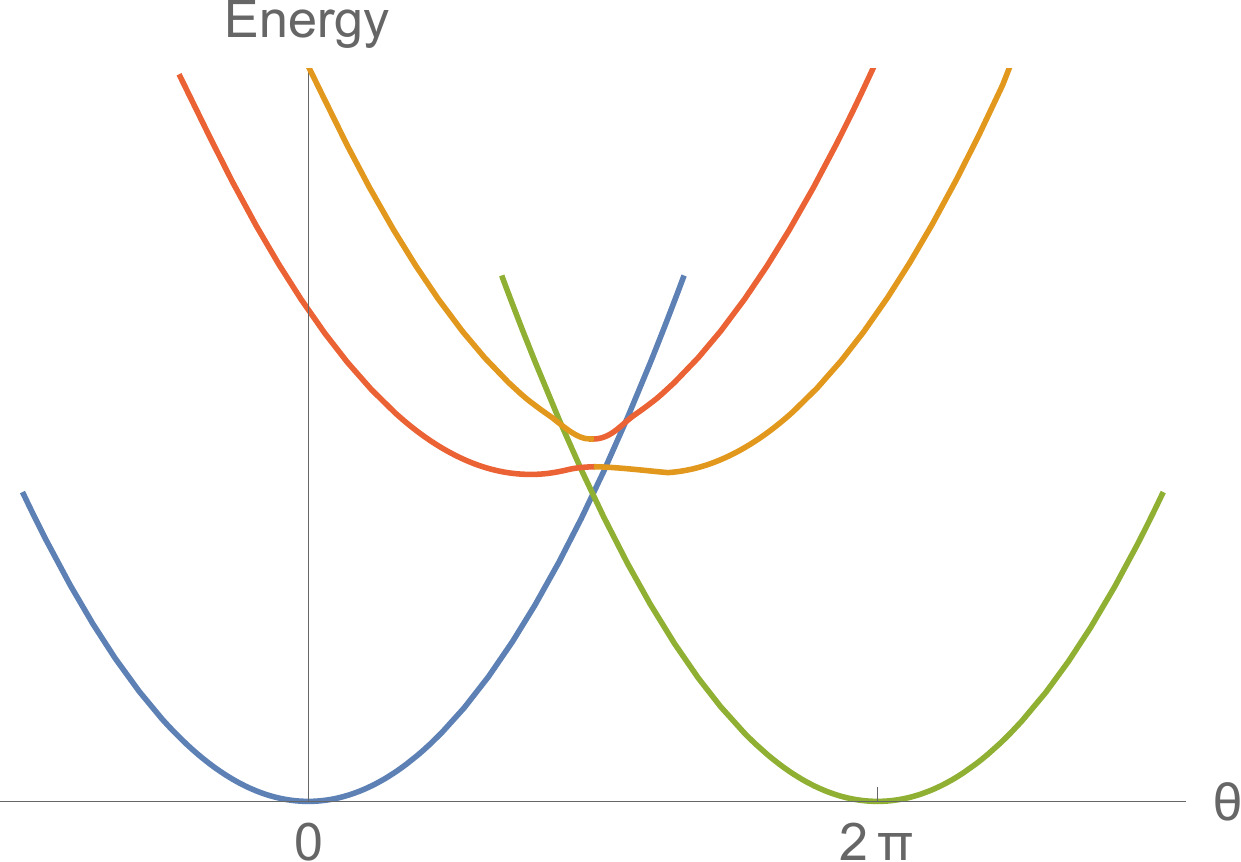}
\label{fig:Bifund_theta02b}
}\end{minipage}
\caption{Two possibilities of the mixing of states due to dynamical bifundamental matter fields when $\theta=\theta_1=\theta_2$ are around $\pi$. }
\label{fig:Bifund_theta02}
\end{figure}
 
If the energy of the mixed states of $(\theta_1/2\pi,1)\oplus (\theta_2/2\pi-1,1)$ and $(\theta_1/2\pi-1,1)\oplus (\theta_2/2\pi,1)$ is lowered by dynamical matter fields  as in Fig.~\ref{fig:Bifund_theta02a}, then the mixed state is selected as the ground state around $\theta=\pi$. In this case, $CP$ (and $(\mathbb{Z}_2)_I$ if exists) need not be broken, but the first-order phase transition happens across the path connecting $\theta=0$ and $\theta=\pi$. 
If the energy of the mixed states of $(\theta/2\pi,1)\oplus (\theta/2\pi-1,1)$ and $(\theta/2\pi-1,1)\oplus (\theta/2\pi,1)$ is lifted as in Fig.~\ref{fig:Bifund_theta02b}, then they drop out from the consideration and there is the first order phase transition from the state with $(\theta/2\pi,1)\oplus (\theta/2\pi,1)$ to the one with $(\theta/2\pi-1,1)\oplus (\theta/2\pi-1,1)$. In this case, $CP$ is spontaneously broken at $\theta=\pi$. 

We can also understand why no phase transition is required when connecting $\theta_1=\theta_2=0$ and $\theta_1=-\theta_2=-\pi$. For instance, let us pick up a path with $\theta_1=-\theta_2$, and denote $\theta'=-\theta_1=\theta_2$. 
By taking the limit $m\to \infty$, we can again consider possible phases using the dual superconductor picture. 
The four-fold degeneracy at $\theta'=\pi$ happens at $m=\infty$, and the condensed charges for those four states are given by 
\bea
&&(-\theta'/2\pi,1)\oplus (\theta'/2\pi,1), \; (-\theta'/2\pi+1,1)\oplus (\theta'/2\pi-1,1), \nonumber\\
&&(-\theta'/2\pi+1,1)\oplus (-\theta'/2\pi,1), \;(-\theta'/2\pi,1)\oplus (\theta'/2\pi-1,1). 
\eea
Figure for the vacuum energy is almost the same with Fig.~\ref{fig:Bifund_theta01} just by replacing the label of charges in a straightforward manner. 

\begin{figure}[t]\centering
\begin{minipage}{.45\textwidth}
\subfloat[$CP$ is unbroken at $\theta'=\pi$]{
\includegraphics[scale=0.5]{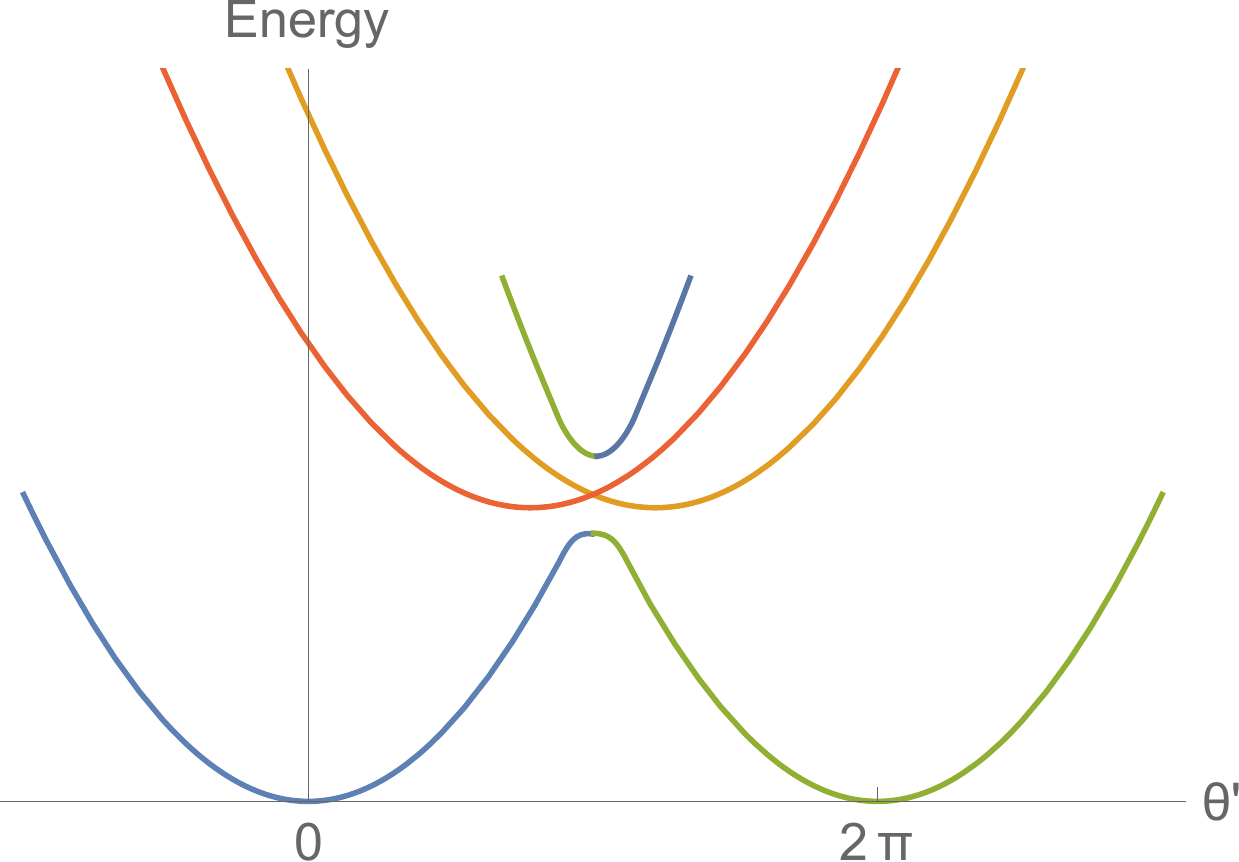}
\label{fig:Bifund_theta03a}
}\end{minipage}\quad
\begin{minipage}{.45\textwidth}
\subfloat[$CP$ is broken at $\theta'=\pi$]{
\includegraphics[scale=0.5]{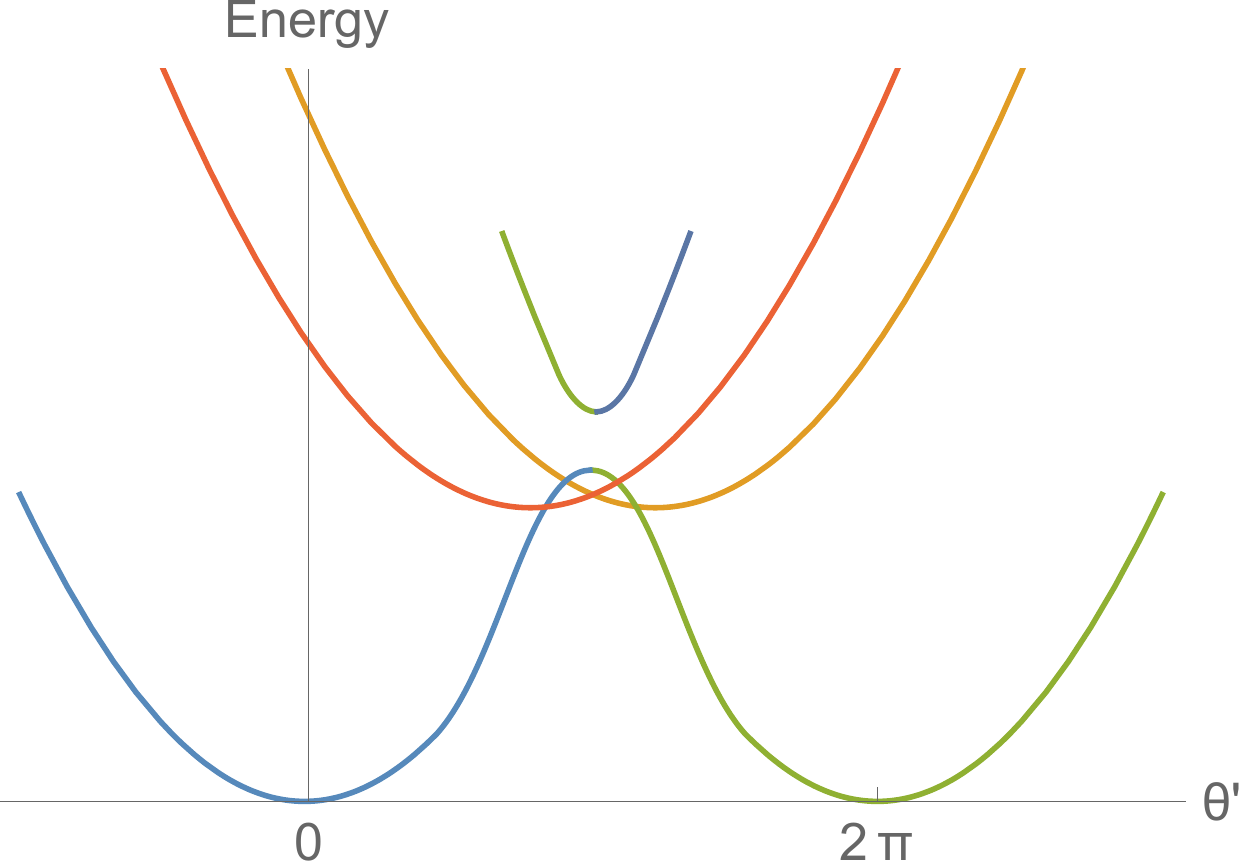}
\label{fig:Bifund_theta03b}
}\end{minipage}
\caption{Two possibilities of the mixing of states due to dynamical bifundamental matter fields when $\theta'=-\theta_1=\theta_2$ are around $\pi$. }
\label{fig:Bifund_theta03}
\end{figure}

Let us turn on dynamical bifundamental fields by making $m$ finite. 
In this case, the states with charges $(-\theta'/2\pi,1)\oplus (\theta'/2\pi,1)$ and  $(-\theta'/2\pi+1,1)\oplus (\theta'/2\pi-1,1)$ can be mixed by dynamical matter fields, while the states with $(-\theta'/2\pi+1,1)\oplus (\theta'/2\pi,1)$ and $(-\theta'/2\pi,1)\oplus (\theta'/2\pi-1,1)$ cannot be mixed. 
Depending on relative energies of those states, we obtain Fig.~\ref{fig:Bifund_theta03} for quasi-vacua of bifundamental gauge theories as a function of $\theta'=-\theta_1=\theta_2$. 
By checking charges of condensed particles, we can notice that Figs.~\ref{fig:Bifund_theta02a} and \ref{fig:Bifund_theta03a} are connected, and $CP$ is unbroken at $\theta_1=\theta_2=\pi$. 
Similarly, Figs.~\ref{fig:Bifund_theta02b} and \ref{fig:Bifund_theta03b} are connected, and $CP$ is spontaneously broken at $\theta_1=\theta_2=\pi$. 
These explain two possible phase boundaries shown in Figs.~\ref{fig:bifund_phase_boundary1} and \ref{fig:bifund_phase_boundary2}, respectively. 

Let us briefly comment on the semiclassical result for this theory on $\mathbb{R}^3\times S^1$ with the double-trace deformation when $g_1^2=g_2^2$ and $\theta_1=\theta_2$~\cite{Shifman:2008ja}. 
Normally the semiclassical computation on the non-Abelian gauge theory breaks down due to the infrared renormalons, and perturbative computations on the small circle compactification gives qualitatively different answers from the nonperturbative results of the theory on $\mathbb{R}^4$. 
By performing a twist with appropriate matter contents and/or some deformations of the theory, the volume dependence of the partition function can become mild so as to evade the phase transition on the compactification radius most probably in the large-$n$ limit~\cite{Shifman:2008ja, Unsal:2007vu, Kovtun:2007py, Unsal:2007jx, Unsal:2008ch, Shifman:2009tp, Argyres:2012ka, Argyres:2012vv, Dunne:2012ae, Dunne:2012zk, Poppitz:2012sw, Anber:2013doa, Cherman:2013yfa, Cherman:2014ofa, Misumi:2014bsa, Misumi:2014jua, Dunne:2016nmc, Cherman:2016hcd, Fujimori:2016ljw, Kozcaz:2016wvy, Sulejmanpasic:2016llc, Yamazaki:2017ulc}. 
According to the computation with this idea for the $SU(n)\times SU(n)$ Yang-Mills theory with ``one'' bifundamental Dirac fermion, the interchange symmetry $(\mathbb{Z}_2)_I$ is unbroken and the first-order phase transition happens at $\theta=\pi$ associated with the spontaneous $CP$ breaking~\cite{Shifman:2008ja}. 
In the language of the dual superconductor model, this suggests that, when $n$ is large with this specific matter content, the states with $(\theta/2\pi,1)\oplus (\theta/2\pi-1,1)$ and $(\theta/2\pi-1,1)\oplus (\theta/2\pi,1)$ are unfavored, and that the first order phase transition occurs at $\theta=\pi$ by jumping from the state with $(\theta/2\pi,1)\oplus (\theta/2\pi,1)$ to the another one with $(\theta/2\pi-1,1)\oplus (\theta/2\pi-1,1)$. 
Let us also comment that this unbroken $(\mathbb{Z}_2)_I$ is the essential ingredient for the orbifold equivalence of this theory, and the question whether it is broken or not on $\mathbb{R}^4$ is not yet settled~\cite{Kovtun:2003hr, Kovtun:2004bz, Kovtun:2005kh, Armoni:2005wta}. 
We cannot answer this question only from our analysis, but let us comment that both phase diagrams shown in Fig.~\ref{fig:bifund_phase_boundary} are consistent with the unbroken $(\mathbb{Z}_2)_I$ symmetry along the line $\theta_1=\theta_2$ when $g_1^2=g_2^2$. 
 
We have so far explained how $CP$ is spontaneously broken at $\theta=\pi$ for bifundamental theories, but our result suggests that it needs not happen if $n=2$. 
We close this section by observing why $n=2$ can be special. 
Let us consider the case with $\theta=\theta_1=\theta_2$ for example, then above conclusion comes from the fact that the states with condensed particles $(\theta/2\pi,1)\oplus (\theta/2\pi,1)$ and $(\theta/2\pi-1,1)\oplus (\theta/2\pi-1,1)$ mod $n$ cannot be mixed.  
This is because the difference of these charges of condensed particles is $(1,0)\oplus (1,0)$ mod $n$, while the charge of dynamical bifundamental matters is $(1,0)\oplus (-1,0)$ mod $n$. These two are different for $n\ge 3$, but they are the same at $n=2$. 
Therefore, for $n=2$, these two states can also be mixed by dynamical bifundamental fields, and thus we need no first-order phase transition lines that separate $\theta=0$ and $\theta=\pi$.

\section{Conclusion}\label{sec:conclusion}

We have studied the phase structure for $SU(n)\times SU(n)$ bifundamental gauge theories at finite topological angles by applying consistency for mixed 't Hooft anomalies of $CP$ and center symmetry. 
For the gauge group $SU(n)\times SU(n)$, there are two topological angles $\theta_1$ and $\theta_2$, and there are four $CP$ invariant points, $(\theta_1,\theta_2)=(0,0)$, $(\pi,0)$, $(0,\pi)$, and $(\pi,\pi)$. 
We discuss that there must be a first-order phase transition at $(\theta_1,\theta_2)=(\pi,0)$ and $(0,\pi)$ associated with spontaneous breaking of the $CP$ symmetry, so there is a first-order phase transition line  through these points. 
The global consistency is discussed at $(\theta_1,\theta_2)=(\pi,\pi)$ but there are two different ways to connect $(\theta_1,\theta_2)=(0,0)$ and $(\theta_1,\theta_2)=(\pi,\pi)$ because the point is equivalent to $(\theta_1,\theta_2)=(-\pi,\pi)$. 
We observe for $n\ge 3$ that the vacua at $(\theta_1,\theta_2)=(0,0)$ and $(\theta_1,\theta_2)=(\pi,\pi)$ cannot be continuously connected without breaking the $CP$ symmetry at $(\theta_1,\theta_2)=(\pi,\pi)$, but also that the vacua at $(\theta_1,\theta_2)=(0,0)$ and $(\theta_1,\theta_2)=(-\pi,\pi)$ can without breaking any symmetries. 
We propose phase diagrams in the $\theta_1$-$\theta_2$ plane that are consistent with these constraints. 
To understand it better, we give a heuristic interpretation of the result based on the dual superconductor model of confinement and the role of dynamical bifundamental fields is clarified. 

The $SU(n)\times SU(n)$ gauge theory with one bifundamental Dirac fermion is a daughter theory of the orbifold equivalence with $\mathcal{N}=1$ supersymmetric $SU(2n)$ Yang-Mills theory in the planar limit at least diagrammatically, and its nonperturbative equivalence is still in question. 
For the nonperturbative equivalence, the interchange symmetry $(\mathbb{Z}_2)_I$ must be unbroken, and we need further investigation for the (non)equivalence. 
Our constraint does not relate the center $\mathbb{Z}_n$ symmetry with the $(\mathbb{Z}_2)_I$ symmetry, so we need more detailed knowledge on dynamics. We point out that the phase diagrams proposed in this paper is consistent with this $(\mathbb{Z}_2)_I$ symmetry when $\theta_1=\theta_2$. 
In order to get microscopic details, numerical simulation is an important subject to study the nonperturbative dynamics. It, however, suffers from the sign problem at finite topological angles, so the technique to cure the sign problem must be further developed, such as Lefschetz-thimbles~\cite{Cristoforetti:2012su, Cristoforetti:2013wha, Fujii:2013sra, Tanizaki:2014xba,Tanizaki:2014tua, Kanazawa:2014qma, Tanizaki:2015pua, DiRenzo:2015foa, Fukushima:2015qza,Tanizaki:2015rda, Fujii:2015bua, Alexandru:2015sua, Alexandru:2016gsd, Tanizaki:2016xcu}, complex Langevin method~\cite{Aarts:2009uq,Aarts:2011ax,Aarts:2014nxa, Nishimura:2015pba,Tsutsui:2015tua,Hayata:2015lzj,Nagata:2016vkn,Salcedo:2016kyy}, etc., for this purpose.  
Careful treatment of the cancellation of these signs is crucial to obtain the physics at $\theta=\pi$ correctly, because the drastic difference between $\theta=0$ and $\theta=\pi$ originates from different interference of the microscopic dynamics in various topological sectors~\cite{Unsal:2012zj}. 


\acknowledgments
The authors thank Hiromichi Nishimura and Zohar Komargodski for fruitful discussions. Y.~T. also thanks Takuya Okuda, Takuya Nishimura, Yoshimasa Hidaka, Yoshio Kikukawa, and Masato Taki for having a study group on higher-form symmetries at Komaba two years ago and it becomes helpful in this study.  Y.~T. is financially supported by RIKEN special postdoctoral program. Y.~K. is supported by the Grants-in-Aid for JSPS fellows (Grant No.15J01626).

\appendix

\section{Quick review on topological field theories}\label{sec:tft}

This section is a minimal reminder for computing the $BF$-type topological field theories, which is necessary in the main text of this paper. 
The basic object is the action 
\be
S={\im \over 2\pi}\int_{X^d} B\diff A^{(i)}, 
\ee
where $X^d$ is a closed, oriented, $d$-dimensional manifold, $A^{(i)}$ is a $U(1)$ $i$-form gauge field, and $B$ is an $\mathbb{R}$-valued $(d-i-1)$-form field. 
The requirement of $U(1)$ $n$-form gauge fields is that $\int_S \diff A^{(i)}\in 2\pi\im \mathbb{Z}$ for any $[S]\in H_{(i+1)}(X,\mathbb{Z})$.  In other words, $\diff A^{(i)}/2\pi \im\in H^{(i+1)}(X,\mathbb{Z})$. 
To understand how the computation goes, let us compute 
\be
\int \Diff A^{(i)} \exp\left({\im\over 2\pi}\int_{X} B\diff A^{(i)}\right). 
\ee
Since $\diff A^{(i)}/2\pi\im \in H^{i+1}(X,\mathbb{Z})$, we can decompose $\diff A^{(i)}$ as 
\be
\diff A^{(i)}=\diff \phi^{(i)}+2\pi \im\sum_{k} n_k \delta(\mathcal{J}_k), 
\ee 
where $\phi^{(i)}$ is a (globally-defined) $i$-form, $n_k$ are integers, $\mathcal{J}_k$ are generators of $H_{d-(i+1)}(X,\mathbb{Z})$, and $\delta(\mathcal{J}_k)$ are their delta-functional forms, i.e. their Poincar\'e duals. 
Therefore, 
\bea
\int \Diff A^{(i)} \exp\left({\im\over 2\pi}\int_{X} B\diff A^{(i)}\right)
&=&\int \Diff \phi^{(i)} \exp\left({\im\over 2\pi}\int_{X} B\diff \phi^{(i)}\right) \prod_k \left\{\sum_{n_k\in\mathbb{Z}}\mathrm{e}^{n_k\int_{\mathcal{J}_k} B}\right\}\nonumber\\
&=&\delta\left({B\over 2\pi\im} \in H^{d-i-1}(X,\mathbb{Z})\right). 
\eea
The last formal expression with the delta function means that this functional integral does not vanish only if $B/2\pi\im$ is an element of the integer-valued cohomology: 
The integration over $\phi^{(i)}$ requires that $\diff B=0$, so $B/2\pi\im\in  H^{d-i-1}(X,\mathbb{R})$. If $\int_{\mathcal{J}_k}B\not\in 2\pi \im \mathbb{Z}$, the summation over $n$ vanishes, and thus we get the result. 

The relevant $4$-dimensional topological field theory in this note is 
\be
S={\im p\over 4\pi n}\int_X \diff A\wedge \diff A, 
\ee
where $X$ is a closed, oriented $4$-manifold, and $A$ is a $U(1)$ $1$-form gauge field. For generic $4$-manifold, $S\in 2\pi \im {p\over 2n}\mathbb{Z}$. To see this, we can write this action as 
\be
S={2\pi \im p\over 2n}\int_X {\diff A \over 2\pi}\wedge {\diff A\over 2\pi}={2\pi p\over 2n}\sum_k n_n\int_{\mathcal{J}_k}{\diff A\over 2\pi}\in {2\pi \im p\over 2n}\mathbb{Z}. 
\ee
Especially when $X$ is a spin manifold, we can use the index theorem to state that the Chern character $\int \exp({\diff A\over 2\pi})\in\mathbb{Z}$, and thus 
\be
S={2\pi \im p\over n}\left({1\over 2}\int_X {\diff A \over 2\pi}\wedge {\diff A\over 2\pi}\right)={2\pi \im p\over n}\int_X \mathrm{e}^{\diff A/2\pi}\in {2\pi\im p\over n}\mathbb{Z}.
\ee
Therefore, $p$ must be identified with mod $2n$ if $X$ is a non-spin manifold, but $p$ must be identified with mod $n$ for spin manifolds. 
Let $X$ be spin and $p\in\mathbb{Z}$, then the theory has the $U(1)$ $1$-form gauge symmetry under 
\be
A\mapsto A-n\lambda, 
\ee
where $\lambda$ is also a $U(1)$ $1$-form gauge field. Indeed, the change of the action under this transformation is 
\be
\Delta S=2\pi \im p \left(-\int{\diff \lambda\over 2\pi}\wedge {\diff A \over 2\pi}+{n\over 2}\int{\diff \lambda\over 2\pi}\wedge {\diff \lambda \over 2\pi}\right). 
\ee
Each term inside the parenthesis gives an integer, and thus $\mathrm{e}^{\Delta S}=1$ when $p\in\mathbb{Z}$. 

Using the auxiliary $\mathbb{R}$-valued $2$-form field $F$ (magnetic field strength) and $U(1)$ $2$-form gauge field $B$, we can dualize this theory as 
\be
S={\im\over 2\pi}\int F\wedge (\diff A+n B)+{\im np\over 4\pi}\int B\wedge B. 
\ee
Integrating out $F$, $nB=-\diff A$ and we obtain the original action. If we integrate out $A$, $F/2\pi \im\in H^2(X,\mathbb{Z})$ and one can introduce a dual $U(1)$ $1$-form gauge field $A_D$ as $F=\diff A_D$. 
Substitution of the result, we get a different expression for the same topological field theory,
\be
S={\im n\over 2\pi}\int B\wedge \diff A_D+{\im n p\over 4\pi}\int B\wedge B. 
\ee
The fact that $F=\diff A_D$ is an important notice for defining dyonic genuine line operators since $\int_\Sigma F=\int_{\p \Sigma}A_D$ does no longer depend on the choice of the surface $\Sigma$ up to an irrelevant phase in $2\pi\im\mathbb{Z}$. 

\bibliographystyle{utphys}
\bibliography{QFT,lefschetz}
\end{document}